\RequirePackage[]{silence}

   \documentclass[fleqn,usenatbib]{mnras}

\usepackage[T1]{fontenc}

\PassOptionsToPackage{pdfpagelabels=false}{hyperref} 

\usepackage{graphicx}	
\usepackage{amsmath}	
\usepackage{amssymb}	

\usepackage{lscape}
\usepackage{subfigure}
\usepackage{multirow}
\usepackage{url}
\usepackage{threeparttable}
\usepackage{natbib}
\title[WAs in NGC 4051]{Nature of the Warm Absorber Outflow in NGC 4051
}

\author[M.\,Mizumoto and K.\,Ebisawa]{
Misaki Mizumoto$^{1,2}$\thanks{E-mail: mizumoto@astro.isas.jaxa.jp (MM)},
and Ken Ebisawa$^{1,2}$
\\
$^1$Institute of Space and Astronautical Science (ISAS), Japan Aerospace Exploration Agency (JAXA), 3-1-1 Yoshinodai, Chuo-ku,\\
Sagamihara, Kanagawa, 252-5210, Japan\\
$^2$Department of Astronomy, Graduate School of Science, The University of Tokyo, 7-3-1 Hongo, Bunkyo-ku, Tokyo, 113-0033, Japan
}

\date{Accepted XXX. Received YYY; in original form ZZZ}

\pubyear{2016}

\begin{document}
\label{firstpage}
\pagerange{\pageref{firstpage}--\pageref{lastpage}}
\maketitle

\begin{abstract}
The Narrow-line Seyfert 1 galaxy NGC 4051 is known to exhibit significant X-ray spectral/flux variations and have a number of emission/absorption features.
X-ray observations have revealed that
these absorption features are blueshifted, which indicates that
NGC 4051 has warm absorber outflow.
In order to constrain physical parameters of the warm absorber outflow,
we analyze the archival data with the longest exposure taken by XMM-Newton in 2009.
We calculate the root-mean-square (RMS) spectra with the grating spectral resolution for the first time.
The RMS spectra have a sharp peak and several dips, which can be explained by variable absorption features and non-variable emission lines;
a lower-ionized warm absorber (WA1: $\log\xi = 1.5,\, v=-650\,\mathrm{km\,s}^{-1}$) shows large variability, 
whereas higher-ionized warm absorbers (WA2: $\log\xi= 2.5,\,v=-4100\,\mathrm{km\,s}^{-1}$, WA3: $\log\xi= 3.4,\, v=-6100\,\mathrm{km\,s}^{-1}$) show little variability.
WA1 shows the maximum variability at a timescale of $\sim10^4$~s,
suggesting that the absorber locates at $\sim10^3$ times of the Schwarzschild radius.
The depth of the absorption features due to WA1 and the observed soft X-ray flux are anti-correlated in several observational sequences,
which can be explained by variation of partial covering fraction of the double-layer blobs
that are composed of the Compton-thick core and
the ionized layer ($=$WA1).
WA2 and WA3 show little variability and presumably extend uniformly in the line of sight.
The present result shows that NGC 4051 has two types of the warm absorber outflows; 
the static, high-ionized and extended line-driven disk winds, and the variable, low-ionized and clumpy double-layer blobs.
\end{abstract}

\begin{keywords}
galaxies: active -- galaxies: Seyfert -- X-rays: individual: NGC 4051
\end{keywords}



\section{Introduction} \label{sec1}

NGC 4051 is an archetypical narrow-line Seyfert 1 (NLS1) at $z=0.0023$ with a mass of $(1.7\pm0.5)\times10^6\,M_\odot$ \citep{den09}.
High-resolution X-ray spectroscopic observations have detected a lot of blueshifted absorption lines 
in the X-ray energy spectra (see, e.g.~\citealt{col01,kro07,pou11a,pou11b}), 
which indicates presence of the warm absorber outflow with discrete velocity components.

In general, geometry of the warm absorber outflow is difficult to constrain.
We can derive the column density ($N_\mathrm{H}=n\Delta r$) and the ionization degree ($\xi=L/(nr^2)$) from model fitting,
where $n$ is electron number density, $\Delta r$ is thickness of the absorber, $r$ is location of the absorber, and $L$ is the X-ray luminosity.
Because these two equations have three unknown values, 
we cannot solve the problem without some assumptions.
For example, \citet{kin12} derived $n$ of the emission-line region from the ratio of triplet emission lines, and
constrained $n$ of the warm absorber assuming that the emission and absorption features originate in the same region.
\citet{kro07} estimated $n$ from photoionization equilibrium time scales.
\citet{ste09} derived these values from recombination time scales.
In the end, they elicit very different geometry of outflows.

In order to constrain geometry of the warm absorber outflow,
we focus on spectral variability.
If we can investigate spectral variability with a high energy resolution,
we are able to determine which components are variable and which are not.
From characteristic time scales of the spectral variation,
we may constrain location of the multiple absorbers and
disentangle their parameters.
In order to obtain sufficient photon counts to investigate detailed spectral variability,
it is vital that the object is bright enough and has strong absorption features, and that the exposure time is sufficiently long;
NGC 4051 is the best target for our analysis because it satisfies all the requirements above.

In this paper, we discuss spectral variability of NGC 4051 with high energy resolution.
First, we explain the observations and data reduction in Section \ref{sec2}.
In Section \ref{sec3}, we show the data analysis, 
where the observed energy spectrum requires three different warm absorbers.
We see that one warm absorber has a rapid variability, whereas the other absorbers show little variations, from which we constrain location of the absorbers.
We discuss origin of the X-ray spectral variability and geometry of the warm absorber outflows of NGC 4051 in Section \ref{sec4}.
Finally we show our conclusion in Section \ref{sec5}.

\section{Observation}\label{sec2}

The XMM-Newton satellite \citep{jan01} observed NGC 4051 fifteen times during May--June 2009.
Observation IDs, start dates, and exposure times are listed in Table \ref{dataset}.
The total good exposure time is 334~ks.
We used the European Photon Imaging Camera (EPIC)-pn data \citep{str01} in the 0.4--12.0 keV band
and the reflection grating spectrometer (RGS: \citealt{den01}) in the 0.4--2.0 keV band.
EPIC-pn was operated in Small Window mode.
We used the XMM-Newton Software Analysis System (SAS, v.13.5.0) and the latest calibration files as of January 2016.
Their spectra and light curves were extracted with {\tt PATTERN$<=$4}
from the circular regions with 30$^{\prime\prime}$ radius centred on the source, 
whereas background products were extracted from circular regions with 60$^{\prime\prime}$ radius within the same CCD chip.
High background periods were excluded from both EPIC and RGS data, when the EPIC/pn count rate of the 10--12 keV band with {\tt PATTERN$==$0} is higher than 0.4~cts/s.
We used {\tt epatplot} to confirm that the pile-up effect is negligible on all the EPIC-pn dataset.
The RGS data were processed with {\tt rgsproc}.
All the spectral fitting were made with {\tt xspec} v.12.8.2 \citep{arn96}.
In the following, the {\tt xspec} model names used in the spectral analysis are explicitly given,
the errors are quoted at the statistical 90\% level,
and the cosmological parameters are as follows: 
$H_0=70\,\mathrm{km}\,\mathrm{s}^{-1}\,\mathrm{Mpc}^{-1}$,
$\Omega_m=0.27$, and $\Omega_\lambda=0.73$.

\begin{table}
\centering
\caption{
Observation IDs, start dates, and good exposure times after removing high background periods.
  }
\label{dataset}
    \begin{tabular}{lllll}
      \hline
\hline
      Name &  Observation ID & Date & Good exposure\\
      \hline
      Obs1  & 0606320101 & 2009-05-03 & 31.4 ks \\
      Obs2  & 0606320201 & 2009-05-05 & 29.1 ks\\
      Obs3  & 0606320301 & 2009-05-09 & 15.8 ks\\
      Obs4  & 0606320401 & 2009-05-11 & 15.1 ks\\
      Obs5  & 0606321301 & 2009-05-15 & 21.0 ks \\ 
      Obs6  & 0606321401 & 2009-05-17 & 26.5 ks \\
      Obs7  & 0606321501 & 2009-05-19 & 22.5 ks\\     
      Obs8  & 0606321601 & 2009-05-21 & 29.0 ks\\ 
      Obs9  & 0606321701 & 2009-05-27 & 26.8 ks\\ 
      Obs10 & 0606321801 & 2009-05-29 & 14.6 ks\\     
      Obs11 & 0606321901 & 2009-06-02 & 43.7 ks\\
      Obs12 & 0606322001 & 2009-06-04 & 20.4 ks\\
      Obs13 & 0606322101 & 2009-06-08 & 24.6 ks\\
      Obs14 & 0606322201 & 2009-06-10 & 23.5 ks\\
      Obs15 & 0606322301 & 2009-06-16 & 29.4 ks\\  
      \hline
    \end{tabular}
\end{table}

\section{Data analysis} \label{sec3}
\subsection{Spectral fitting}\label{sec3.1}
\begin{figure}
\includegraphics[width=100mm]{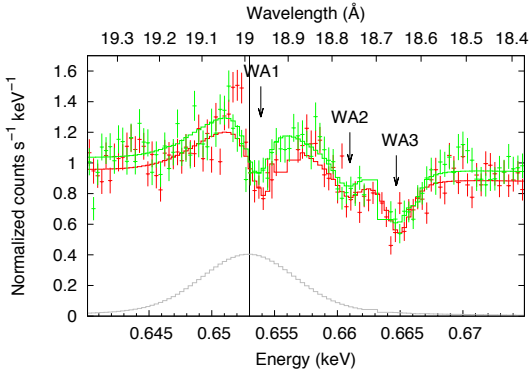}
\caption{The \ion{O}{VIII} Lyman $\alpha$ emission line and three absorption lines fitted with Gaussians.
The red/green show the 1st order of RGS1/RGS2 spectra, respectively.
The grey Gaussian shows the emission line.
The central energy of the emission line, which has no velocity in the AGN frame, is shown by the vertical line, and those of the absorption lines are shown by the arrows. 
}
\label{fig:rgs}
\end{figure}

We stacked all the data and created the time-averaged spectra.
The spectral continuum of the EPIC energy band is explained by a power-law component and a soft excess.
We fitted the soft excess with the multi-colour disk ({\tt diskbb}; \citealt{mit84}),
and fitted the power law component with {\tt cutoffpl}.
The EPIC spectrum has a fluorescent Fe-K line at 6.4~keV,
therefore we added a reflected component from neutral material ({\tt pexmon}; \citealt{nan07}).
{\tt pexmon} is the model combining {\tt pexrav} \citep{mag95} with self-consistently generated Fe and Ni lines, thus
we can constrain the normalization of the reflection component from strength of the Fe K$\alpha$ line.
When we assume that {\tt pexmon} is at rest in the frame of NGC 4051, some residual is seen.
Therefore we set the velocity of {\tt pexmon} free, to find that the velocity is $-1800_{-400}^{+500}$~km~s$^{-1}$.
The column density of the galactic absorption toward this object is $(1.3\pm0.1)\times10^{20}$~cm$^{-2}$ \citep{elv89},
for which we added {\tt tbabs} \citep{wil00}.
As for the photoionized cross-section of {\tt tbabs},
we used the one calculated by \citet{bcmc} and \citet{yan98}.
In order to fit the absorption features seen in the RGS spectra,
we used a warm absorber model via XSTAR Version 2.2.1bn21 \citep{kal04},
assuming the solar abundance and the photon index of the ionizing spectrum to be 2.0. 
We made a grid model by running XSTAR for different values of $\xi$ and $N_\mathrm{H}$;
the $\log\xi$ values were from 0.1 to 5 (erg cm s$^{-1}$), and
the $N_\mathrm{H}$ values were from $5\times10^{20}$ to $5\times10^{24}$ (cm$^{-2}$).
Figure \ref{fig:rgs} shows the \ion{O}{VIII} spectral shape, which has an emission line and three absorption lines.
We fitted the spectral shape with a phenomenological model with powerlaw, one positive Gaussian, and three negative Gaussians.
All of three negative Gaussians, shown by the arrows in Figure \ref{fig:rgs}, are blueshifted; their velocities are 
$-600^{+90}_{-50}$~km~s$^{-1}$, $-4040^{+70}_{-100}$~km~s$^{-1}$, and $-5780\pm80$~km~s$^{-1}$.
This shows that at least three independent warm absorbers are necessary to explain the observed absorption lines.
We call these warm absorbers as WA1, WA2, and WA3, which have different ionization statues and blueshifts.
Emission lines, including radiative recombination continua (RRC), are also seen in the energy spectra, which are also observed by \citet{nuc10}.
We added positive Gaussians to explain these emission features.
Table \ref{emissionline} shows a list of the emission lines.

\begin{table}
\centering
\caption{
Identification of the observed emission lines; theoretically expected centroid energy ($E_\mathrm{exp}$), observed ones ($E_\mathrm{obs}$), and observed equivalent width (EW) are indicated.
  }
\label{emissionline}
  \begin{threeparttable}
    \begin{tabular}{llll}
      \hline
\hline
Line & $E_\mathrm{exp}$\tnote{$\ast$1} & $E_\mathrm{obs}$ & EW\\ 
      \hline
\ion{N}{VI} (f) & 0.420~keV & 0.419~keV & 0.7~eV\\
\ion{O}{VII} (f) &0.560~keV& 0.560~keV & 3.2~eV\\
\ion{O}{VII} (i) &0.569~keV& 0.568~keV & 1.0~eV\\
\ion{O}{VII} (r) & 0.574~keV & 0.573~keV & 2.0~eV\\
\ion{O}{VIII} Ly-$\alpha$ &0.654~keV& 0.653~keV & 8.3~eV\\
\ion{Fe}{XVII} 3s-2p & 0.726~keV & 0.726~keV & 2.2~eV\\
\ion{Ne}{IX} (f) &0.905~keV& 0.903~keV & 4.0~eV\\
\ion{Ne}{IX} (i) &0.915~keV& 0.912~keV & 0.7~eV\\
\hline
RRC of \ion{C}{VI} &0.490~keV& 0.492~keV & 0.5~eV\\
RRC of \ion{O}{VII} &0.739~keV& 0.735~keV & 2.6~eV\\
      \hline
    \end{tabular}
\begin{tablenotes}\footnotesize
\item[$\ast$1] From the CHIANTI database \citep{der01}.
\end{tablenotes}
  \end{threeparttable}
\end{table}

\begin{figure*}
\includegraphics[width=160mm]{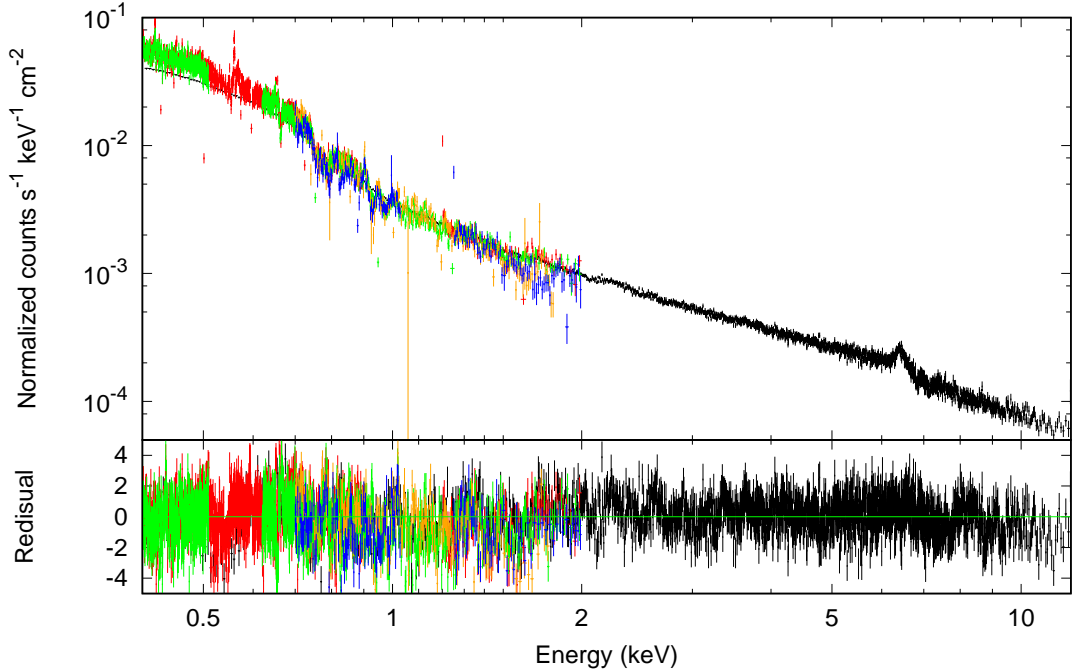}
\caption{Spectral fitting of NGC 4051. The black, red, green, orange, and blue show the EPIC-pn, 1st order of RGS1, 1st order of RGS2, 2nd order of RGS1, and 2nd order of RGS 2 spectra, respectively.
The black line shows the model.
The upper panel shows count plots, removing effect of effective areas (``setplot area'' in xspec), and the lower panel shows the residuals ($\chi$), which mean (data$-$model)/error.
The shown data are more binned than those used in the spectral fitting for clarity.
}
\label{fig:allplot}
\end{figure*}

\begin{figure*}
\includegraphics[width=150mm]{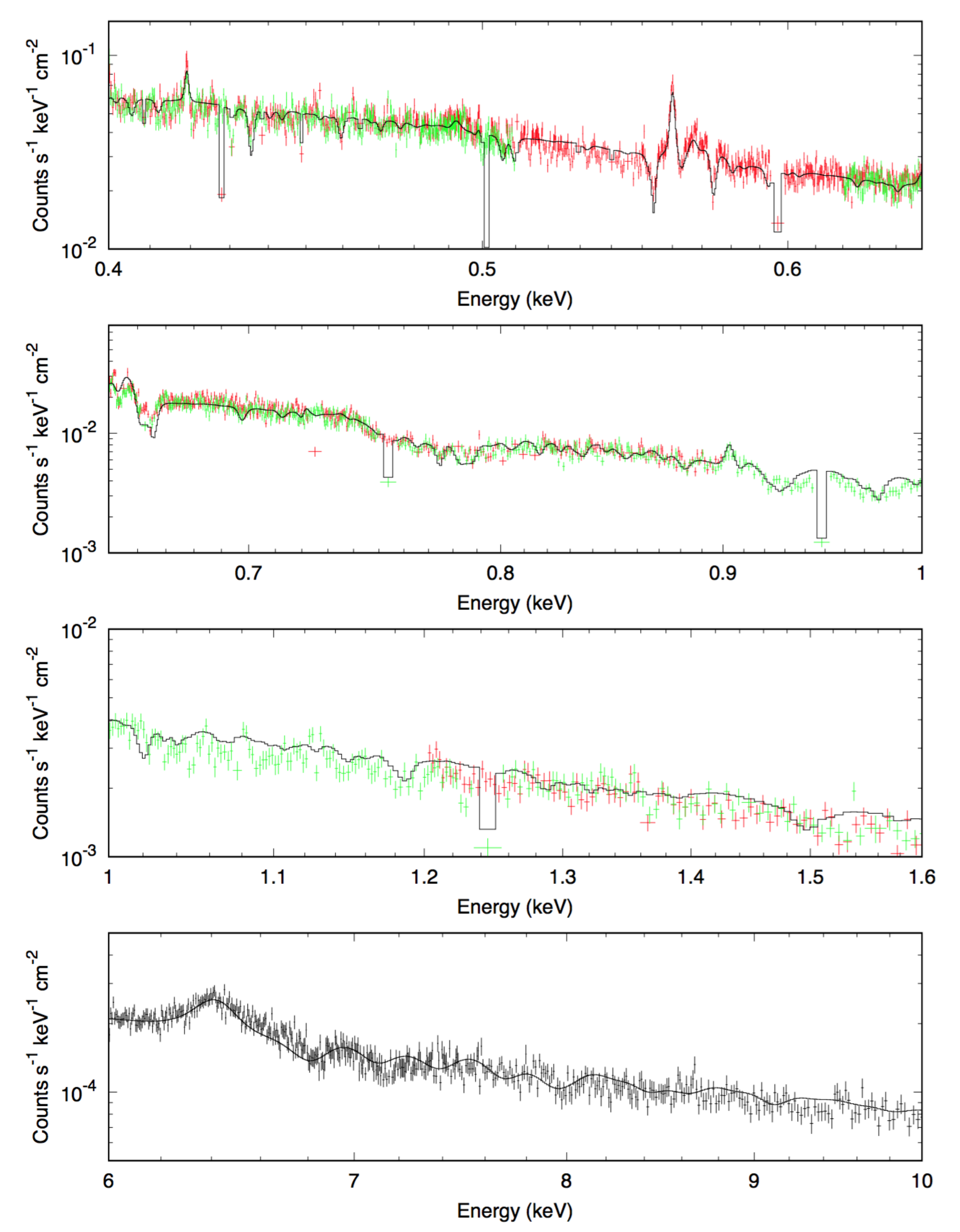}
\caption{Enlargement of the upper panel of Fig.\,\ref{fig:allplot} in order to clarify the emission/absorption line features.
The model lines are those of RGS1 (0.4--0.65 keV), RGS2 (0.65--1.6 keV), and EPIC-pn (6--10 keV).
}
\label{fig:allplot2}
\end{figure*}

\begin{figure}
\includegraphics[width=100mm]{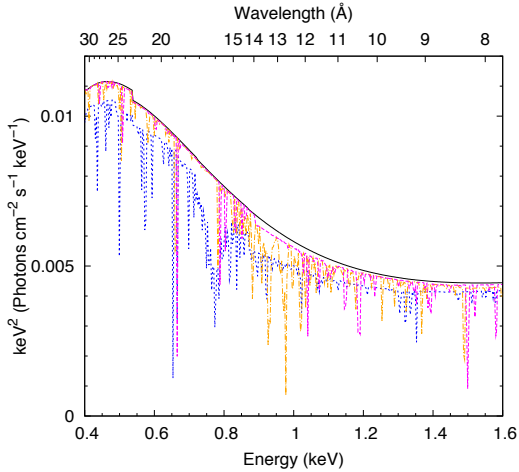}
\caption{Model spectra of three warm absorbers.
The black line shows the continuum ($\mathtt{tbabs}\times(\mathtt{diskbb}+\mathtt{cutoffpl})$), and
the blue dotted, orange dot-dashed, and magenta dashed lines show the continuum with WA1, WA2, and WA3, respectively.
}
\label{fig:eeuf_rgs}
\end{figure}

We simultaneously fitted all the spectra with the following model,
\begin{equation}
\begin{split}
F&=\mathtt{tbabs}\times\{(\mathtt{diskbb}+\mathtt{cutoffpl})\times \mathrm{WA1} \times \mathrm{WA2} \\
&\quad \times \mathrm{WA3} +\mathtt{pexmon}+\mathrm{emission}\:\:\mathrm{lines}\}, \label{eq:eq1}
\end{split}
\end{equation}
and found that this model can almost fully explain the spectra, except two absorption lines.
The \ion{O}{V} 0.554~keV absorption line remains in the spectra, 
thus we added an additional narrow negative Gaussian.
In addition, depth of the \ion{N}{VII} 0.500~keV absorption line seems to be overestimated in the model,
thus we added a positive Gaussian, which might be a real emission line \citep{nuc10}.
Figures \ref{fig:allplot}, \ref{fig:allplot2} and Table \ref{fitting} show the fitting results.
The reduced chi square ($\chi_\nu^2$) is 1.52 for the degree of freedom (dof) of 8928.
Figure \ref{fig:eeuf_rgs} shows the absorption features created by WA1, WA2, and WA3.
We can see that WA1 creates a deep Fe-L unresolved transition array (UTA) feature at $\simeq 0.8$~keV \citep{beh01}.


\begin{table}
 \begin{center}
  \caption{Parameters of spectral fitting
  }\label{fitting}
    \begin{threeparttable}
  \begin{tabular}{lll}
   \hline
\hline
{\tt tbabs} & $N_\mathrm{H}$ (cm$^{-2}$)& $1.3\times10^{20}$ (fix) \tnote{$\ast$1} \\
\hline
{\tt diskbb}& $T_\mathrm{in}$~(eV) &  $155.9\pm0.6$ \\
& Norm.\tnote{$\ast$2} & $2610\pm40$ \\
{\tt cutoffpl} & $\Gamma$ &  $1.581\pm0.006$ \\
& $E_c$~(keV) &  500 (fix) \\
& Norm.\tnote{$\ast$3} & $(3.42\pm0.02)\times10^{-3}$  \\
{\tt pexmon} & $\Omega/2\pi$ & $0.49\pm0.07$\\
& Inclination & $60^\circ$ (fix)\\
& $v$ (km~s$^{-1}$) &  $-1800_{-400}^{+500}$\\
\hline
WA1 & $N_\mathrm{H}$ & $(2.46\pm0.07)\times10^{21}$ \\
& $\log\xi$ & $1.469\pm0.009$\\
& $v$ (km~s$^{-1}$) & $-650\pm20$ \\
WA2 & $N_\mathrm{H}$ & $(8.2_{-0.3}^{+0.4})\times10^{21}$   \\
& $\log\xi$ & $2.512\pm0.013$ \\
& $v$ (km~s$^{-1}$) &  $-4060\pm60$\\
WA3 & $N_\mathrm{H}$ & $(7.1_{-0.9}^{1.1})\times10^{23}$  \\
& $\log\xi$ & $3.383_{-0.006}^{+0.011}$\\
& $v$ (km~s$^{-1}$) &  $-6120\pm20$ \\
\hline
  \end{tabular}
\begin{tablenotes}\footnotesize
    \item[$\ast$1] From \citet{elv89}
    \item[$\ast$2] $\left(\frac{r_\mathrm{in}/\mathrm{km}}{D/10\mathrm{kpc}}\right)^2\cos i$, where $r_\mathrm{in}$ is the inner radius, $D$ is the distance, and $i$ is the inclination angle.
    \item[$\ast$3] Photons~keV~$^{-1}$~cm$^{-2}$~s$^{-1}$ at 1~keV
    \end{tablenotes}
  \end{threeparttable}
  \end{center}
\end{table}
%

\subsection{RMS spectra} \label{sec3.2}
\subsubsection{Calculation of RMS}
In order to evaluate spectral variability, 
we calculated root-mean-square (RMS) of the data.
We adopt the fractional variability amplitude ($F_\mathrm{var}$) and the point-to-point fractional variability ($F_\mathrm{pp}$), which are defined as 
\begin{equation}
F_\mathrm{var}=\frac{1}{\langle X \rangle}\sqrt{S^2-\langle\sigma^2_\mathrm{err} \rangle}
\end{equation}
and
\begin{equation}
F_\mathrm{pp}=\frac{1}{\langle X \rangle}\sqrt{\frac{1}{2(N-1)}\sum_{i=1}^{N-1}(X_{i+1}-X_i)^2 - \langle\sigma^2_\mathrm{err} \rangle} ,
\end{equation}
where $X_i$ is the count for the $i$-th of $N$ bins, 
$\langle X \rangle$ is the mean count rate, 
$S^2$ is the variance of the light curve, and 
$\langle\sigma^2_\mathrm{err} \rangle$ is the mean error squared \citep{ede02}.
The error on $F_\mathrm{var,pp}$ is given as
\begin{equation}
\sigma_{F_\mathrm{var,pp}}=\frac{1}{F_\mathrm{var,pp}}\sqrt{\frac{1}{2N}}\frac{S^2}{\langle X \rangle^2}.
\end{equation}
$F_\mathrm{var}$ shows the long-timescale variability across the whole observation period ($\sim10^5$~s in this dataset), whereas
$F_\mathrm{pp}$ extracts the variability at a given time scale.
See the Appendix of \citet{ede02} how to derive these equations.

\begin{figure}
\centering
\includegraphics[width=100mm]{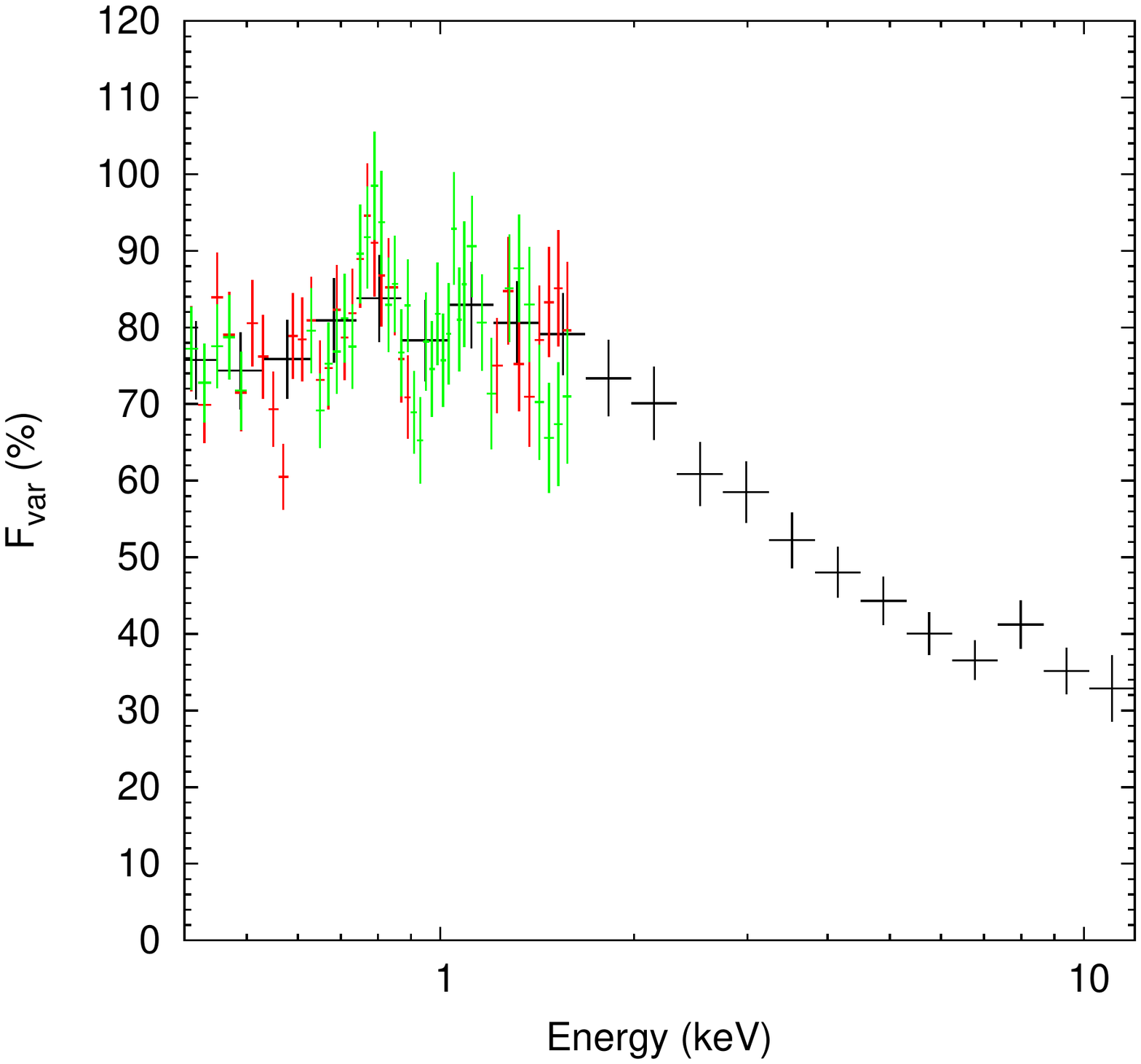}\\
\includegraphics[width=100mm]{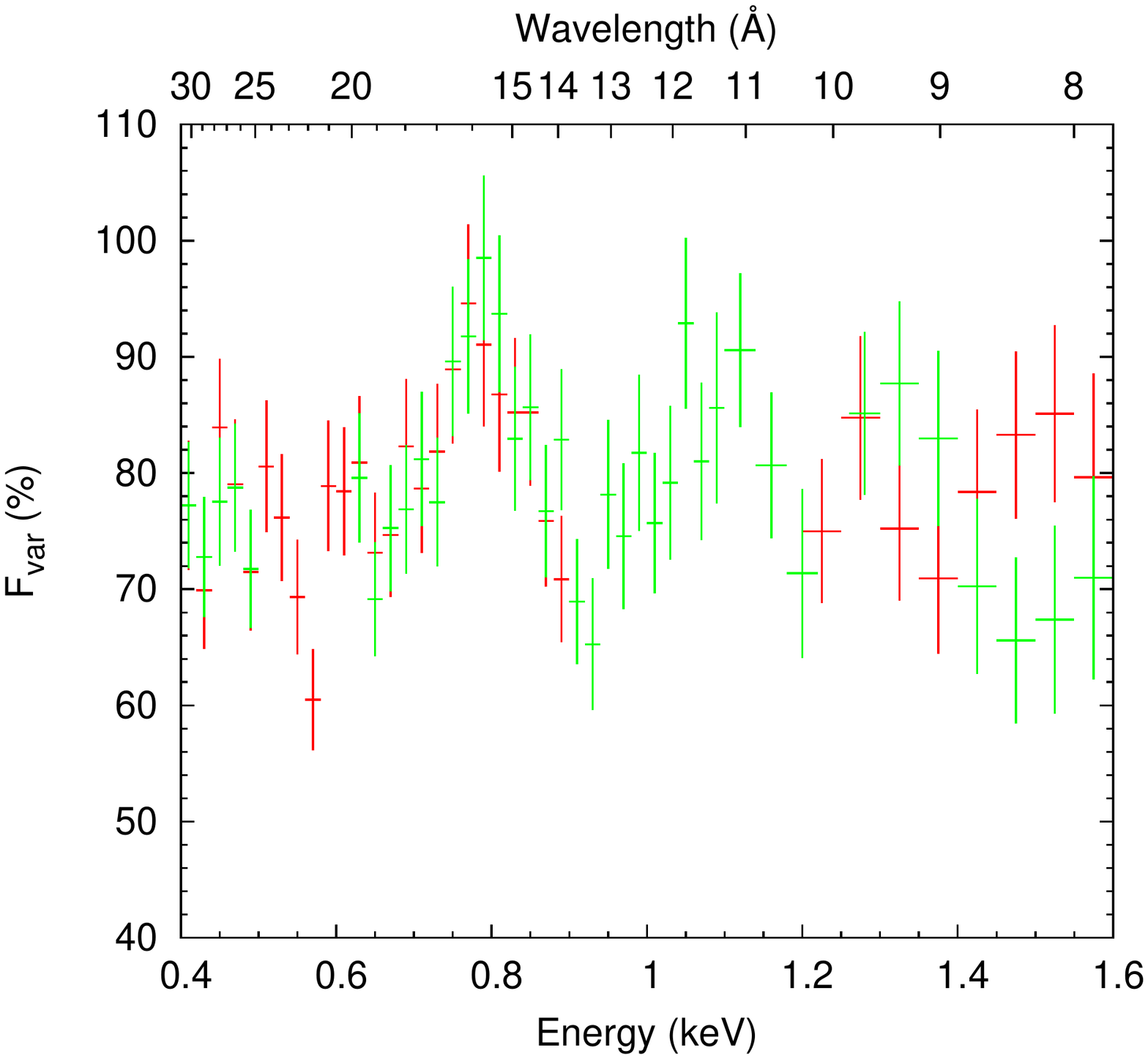}
\caption{RMS spectra of EPIC-pn (black) and 1st order of RGS1/2 (red/green), respectively, with a bin-width of 5000~s.
The lower panel is an enlargement of the upper panel (only RGS data).
}
\label{fig:RMS_all}
\end{figure}

We calculated $F_\mathrm{var}$ and $F_\mathrm{pp}$ using all the fifteen observations.
When calculating RMS of the RGS data,
we used only the 1st order of the 0.4--1.6~keV energy band to maximize photon statistics.
Figures \ref{fig:RMS_all} shows the $F_\mathrm{var}$ spectra with a time bin-width of 5000~s for the entire observation period ($N=111$).
We can see several important features in the RMS spectra.
First, the bin at $\simeq6.4$~keV drops,
which is considered as due to the fluorescent Fe-K line that has little variability \citep{pon06,ter09}.
The RMS spectrum has a peak in the soft energy band, and the value gradually decreases toward higher energies, which is common in several Seyfert 1 galaxies (\citealt{ter09} and referenced therein).
We can see more detailed features in the RGS data;
in particular, it has a strong peak at 0.8~keV, and sharp drops at 0.55~keV and 0.9~keV.

\subsubsection{Absorption/Emission features}
First, we focus on the RMS peak at $\simeq0.8$~keV.
This energy corresponds to that of the Fe-L UTA.
When depth of the absorption feature varies,
such a RMS peak is created.
When only $N_\mathrm{H}$ of the absorber varies,
RMS is calculated to be 
\begin{equation}
F_\mathrm{var}=\left( N_{\mathrm{H}}\right)_\mathrm{var}\cdot\frac{ \sigma(E) \langle  N_\mathrm{H} \rangle}{1-\sigma(E)\langle N_\mathrm{H} \rangle}, \label{eq:absFvar}
\end{equation}
where $\sigma(E)$ is the photoionized cross section, 
$\langle N_\mathrm{H} \rangle$ is the average value of $N_\mathrm{H}$, and 
$\left(N_\mathrm{H}\right)_\mathrm{var}$ is the variability amplitude of $N_\mathrm{H}$  (Appendix \ref{sec:A1}).
In the optically-thin regime, 
variation of $N_\mathrm{H}$ is equivalent to variation of the partial covering fraction as follows:
\begin{eqnarray}
W&=& \exp[-\sigma N_\mathrm{H}] \nonumber \\
&\simeq& 1-\sigma N_\mathrm{H}\nonumber \\
&=&1-\alpha\sigma N_\mathrm{H}^\mathrm{fixed}\nonumber \\
&=&1-\alpha+\alpha(1-\sigma N_\mathrm{H}^\mathrm{fixed})\nonumber \\
&\simeq&1-\alpha+\alpha\exp[-\sigma N_\mathrm{H}^\mathrm{fixed}] \nonumber \\
&=&1-\alpha+\alpha W^\mathrm{fixed}, \label{eq:VDPC}
\end{eqnarray}
where  $\alpha$ is the partial covering fraction and
$N_\mathrm{H}^\mathrm{fixed}$ is the full-covering column density
(see Eq.\,5 of \citealt{miz14}).
When the partial covering fraction varies,
RMS is calculated to be 
\begin{equation}
F_\mathrm{var}=(\alpha)_\mathrm{var}\cdot\frac{ \left(1-\exp\left[-\sigma(E) N_\mathrm{H}^\mathrm{fixed}\right]\right) \langle \alpha \rangle}{1-\left(1-\exp\left[-\sigma(E) N_\mathrm{H}^\mathrm{fixed}\right]\right) \langle \alpha \rangle}, \label{eq:absFvar2}
\end{equation}
(Appendix \ref{sec:A2}).
Eq.\,(\ref{eq:absFvar}) and Eq.\,(\ref{eq:absFvar2}) are mathematically equivalent;
both equations monotonically increase with $\sigma(E)$, which means that
RMS peaks appear at the energy band where the absorption features are deeper than the adjacent bands (see Figure 10 of \citealt{yam16}).

We calculated simulated RMS spectra in order to illustrate effects of the warm absorber variations in the RMS spectra.
The method is as follows.
(1) We set a range of the variable column densities as [$\langle N_{\mathrm{H},k} \rangle -\Delta N_{\mathrm{H},k}$ : $\langle  N_{\mathrm{H},k} \rangle +\Delta N_{\mathrm{H},k}$] for each WA$k$ ($k=1,2,3$), 
or, equivalently a range of the variable partial covering fractions as [$\langle \alpha_k \rangle -\Delta \alpha_k$ : $\langle \alpha_k \rangle +\Delta \alpha_k$].
(2) We substituted $N_{\mathrm{H},k}$ or $\alpha_k$ of the $k$-th absorber random values within the covering range, and
created simulated energy spectra using {\tt fakeit} command in {\tt xspec}.
(3) We repeated the procedure 100 times, and got 100 simulated energy spectra for each $k$.
(4) We calculated the simulated RMS spectra from the 100 simulated energy spectra.

\begin{figure}
\centering
\includegraphics[width=100mm]{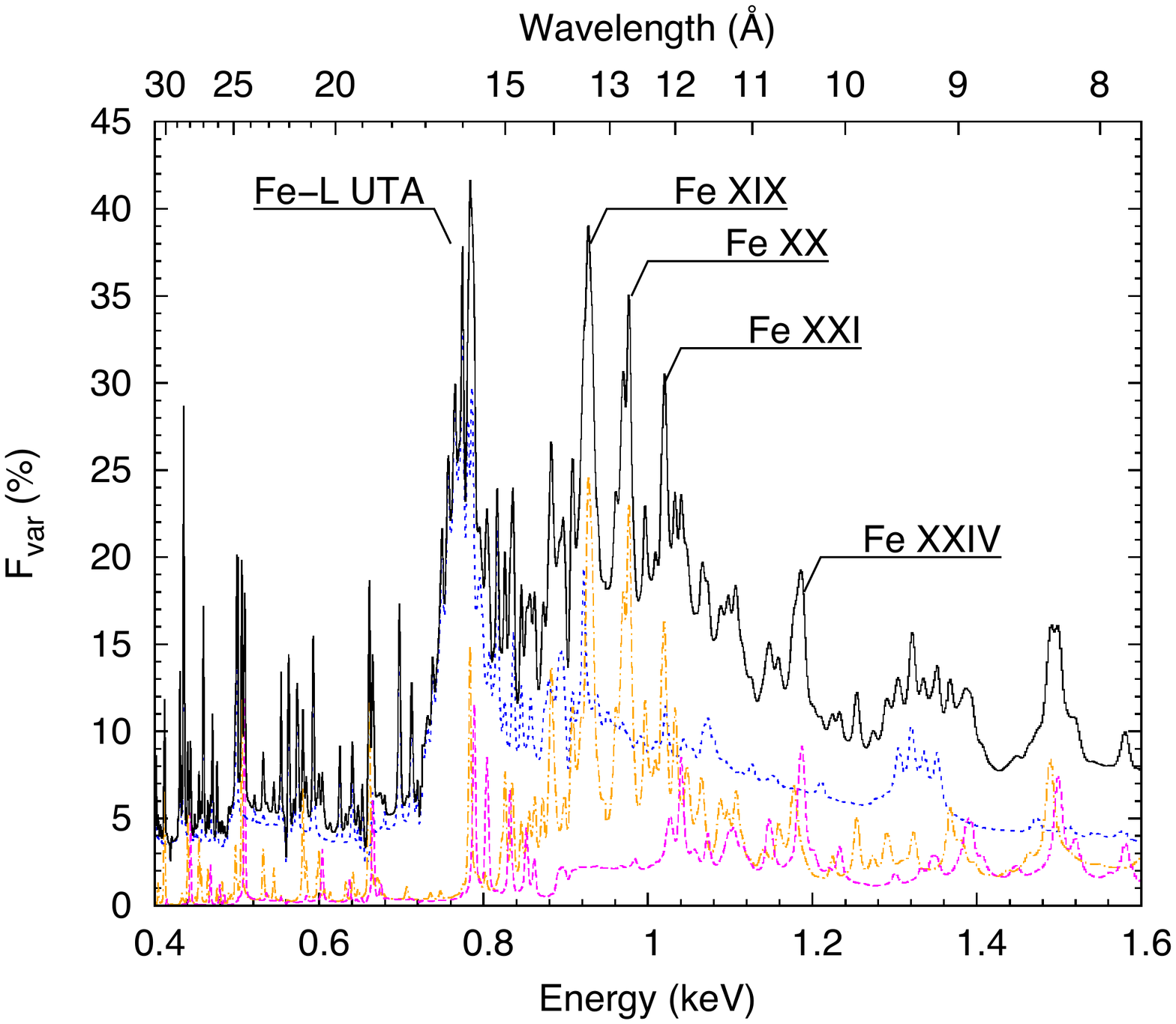}\\
\includegraphics[width=100mm]{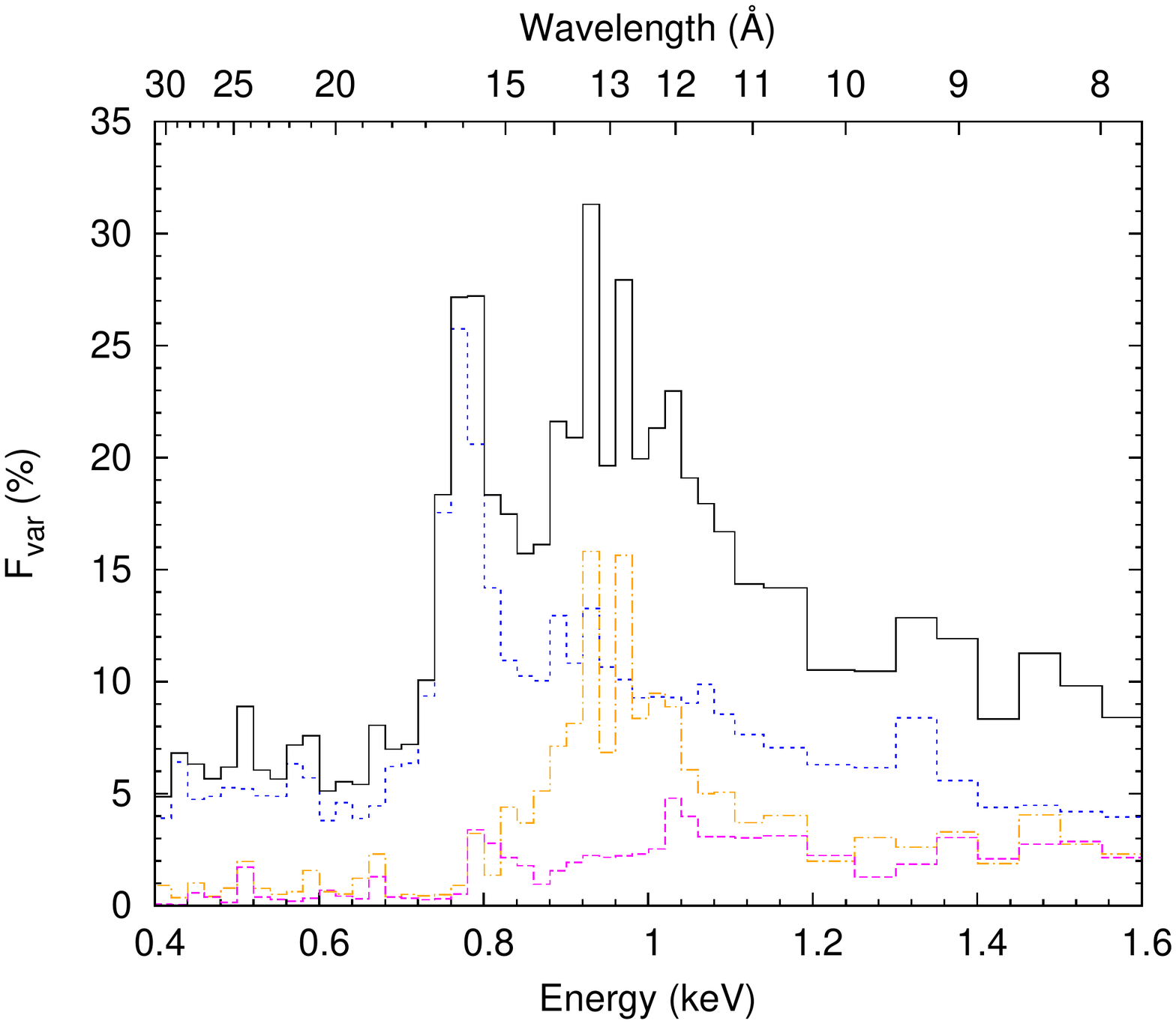}
\caption{Simulated RMS spectra when either column densities or partial covering fractions of the warm absorbers vary. 
The blue dotted, orange dot-dashed, and magenta dashed lines show the effect of WA1, WA2, and WA3, respectively, and
the black line is the sum of them.
The lower panel is the same as the upper one, but binned with the bin-width of Figure \ref{fig:RMS_all} bottom.
}
\label{fig:RMS_abs}
\end{figure}

Figure \ref{fig:RMS_abs} show the simulated RMS spectra.
The blue/orange/magenta lines show the RMS spectra 
when the parameter of WA1/WA2/WA3 varies, respectively, at the same amounts.
The black line is the sum of them.
We set $\Delta N_{\mathrm{H},k}/\langle N_{\mathrm{H},k} \rangle$ or $\Delta \alpha_k / \langle \alpha_k \rangle$ as 100\%,
i.e. $\left(N_{\mathrm{H}}\right)_\mathrm{var}=(\alpha)_\mathrm{var}=100/\sqrt{3}=57.7\%$, for each $k$  (Appendix \ref{sec:A1}).
We can see that the RMS peak at $\sim0.8$~keV is explained by the Fe-L UTA feature of WA1.
In addition, other peaks are seen at the $0.9-1.0$~keV band from WA2 and WA3.

\begin{figure}
\centering
\includegraphics[width=100mm]{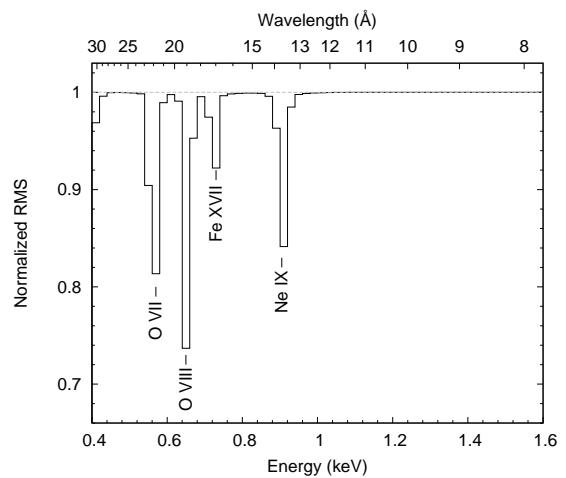}
\caption{Simulated RMS spectra when emission lines are not variable.
}
\label{fig:RMS_line}
\end{figure}

Next, we focus on the RMS dips at 0.55~keV and 0.9~keV.
When emission lines have little variations whereas the entire spectrum varies,
the variability amplitude at the lines are less than those of the adjacent ranges.
When the adjacent range varies with a variability amplitude of $F_\mathrm{var,adj}$,
the variability amplitude of the non-variable emission lines ($F_\mathrm{var,line}$) is calculated as
\begin{equation}
F_\mathrm{var,line}=F_\mathrm{var,adj}\cdot\frac{1}{1+x}, \label{eq:varline}
\end{equation}
where $x$ is the intensity ratio of the line to the adjacent continuum (\ref{sec:A3}).
This means that RMS spectra show drops at the non-variable emission lines.
In order to evaluate effect of the emission lines,
we simulated the RMS spectra, where
only normalization of the continuum is varied.
Figure \ref{fig:RMS_line} shows the effect of the non-variable lines, that is $1/(1+x)$ of Eq.~(\ref{eq:varline}).
We can see that these lines appear as significant drops in the RMS spectra,
if only the continuum is variable.

\begin{figure}
\centering
\includegraphics[width=100mm]{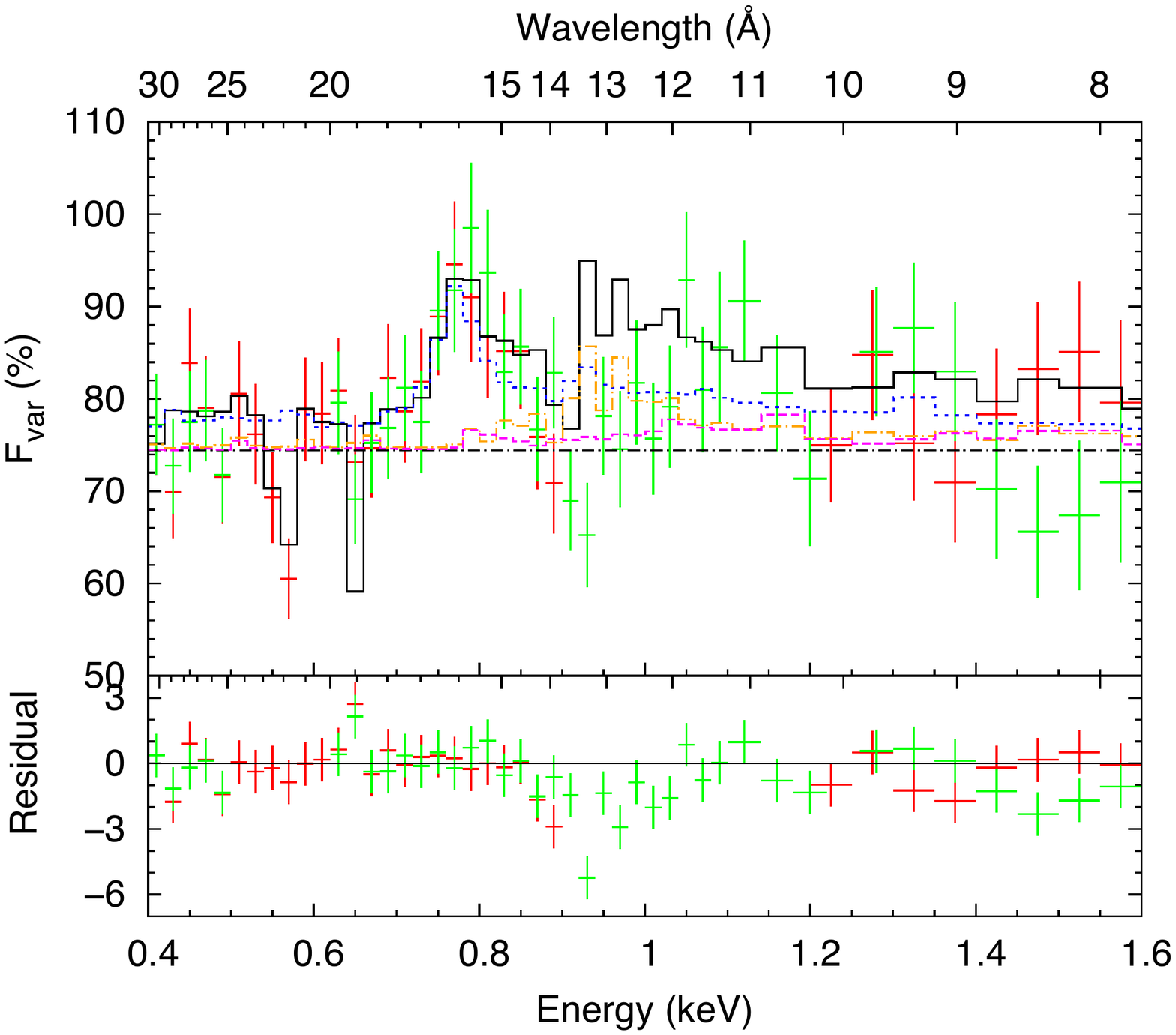}
\caption{Fitting of the $F_\mathrm{var}$ spectra with the simulated ones
when all the warm absorbers have the same variability amplitude.
The red/green bins show the data of 1st order of RGS1/RGS2.
The types of the model lines are same as Figure \ref{fig:RMS_abs}, and
the black dashed line shows the continuum level.
}
\label{fig:RMS_nofitting}
\end{figure}

We compared the observed $F_\mathrm{var}$ spectra with the simulated ones.
If all the warm absorbers had the same variability amplitudes,
a significant peak at the $0.9-1.0$~keV band from WA2 and WA3 would be seen (Figure \ref{fig:RMS_nofitting}).
However, the $F_\mathrm{var}$ spectra have no peak at the energy band,
thus we propose that only WA1 has a large variability,
whereas WA2 and WA3 have little variability.

\begin{figure}
\centering
\includegraphics[width=100mm]{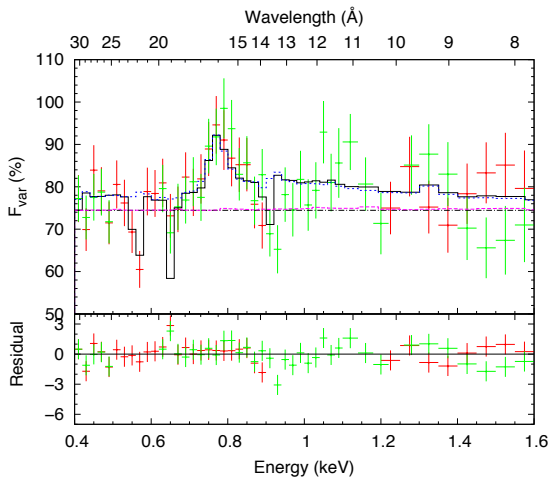}
\caption{Same as Figure \ref{fig:RMS_nofitting}, but
the variability amplitude of each warm absorber is set to be free.
}
\label{fig:RMS_fitting}
\end{figure}

Figure \ref{fig:RMS_fitting} and table \ref{tab:RMS_fitting} show the fitting results,
where parameters are the fractional variability of three warm absorbers and the continuum normalization.
All the features in the $F_\mathrm{var}$ spectra are explained by the model fitting ($\chi_\nu^2=0.99$ for $\mathrm{dof}=68$).
In particular, the RMS dips at $\simeq0.55$~keV and $\simeq0.9$~keV are explained well by the
\ion{O}{VII} (f) and \ion{Ne}{IX} (f) emission lines.
WA1 varies significantly as much as $42\pm8$~\%, 
whereas WA2 varies little ($<5$~\%), and
variation of WA3 cannot be constrained ($ < 49$~\%).
We notice that the continuum level shows a large variability,
which means that the observed soft X-ray flux varies due to other mechanisms.

\begin{table}
\centering
\caption{
Parameters of the $F_\mathrm{var}$ spectral fitting
  }
\label{tab:RMS_fitting}
    \begin{tabular}{ll}
      \hline
\hline
& Fractional variability amplitude \\
\hline
WA1 & $42\pm8$~\% \\
WA2 & $0^{+5}$~\%  \\
WA3 & $9\pm40$~\% \\
Continuum & $74.5\pm1.3$~\% \\
\hline
$\chi_\nu^2$ (dof) & 0.99 (68) \\
\hline
    \end{tabular}
\end{table}

\begin{figure*}
\includegraphics[width=170mm]{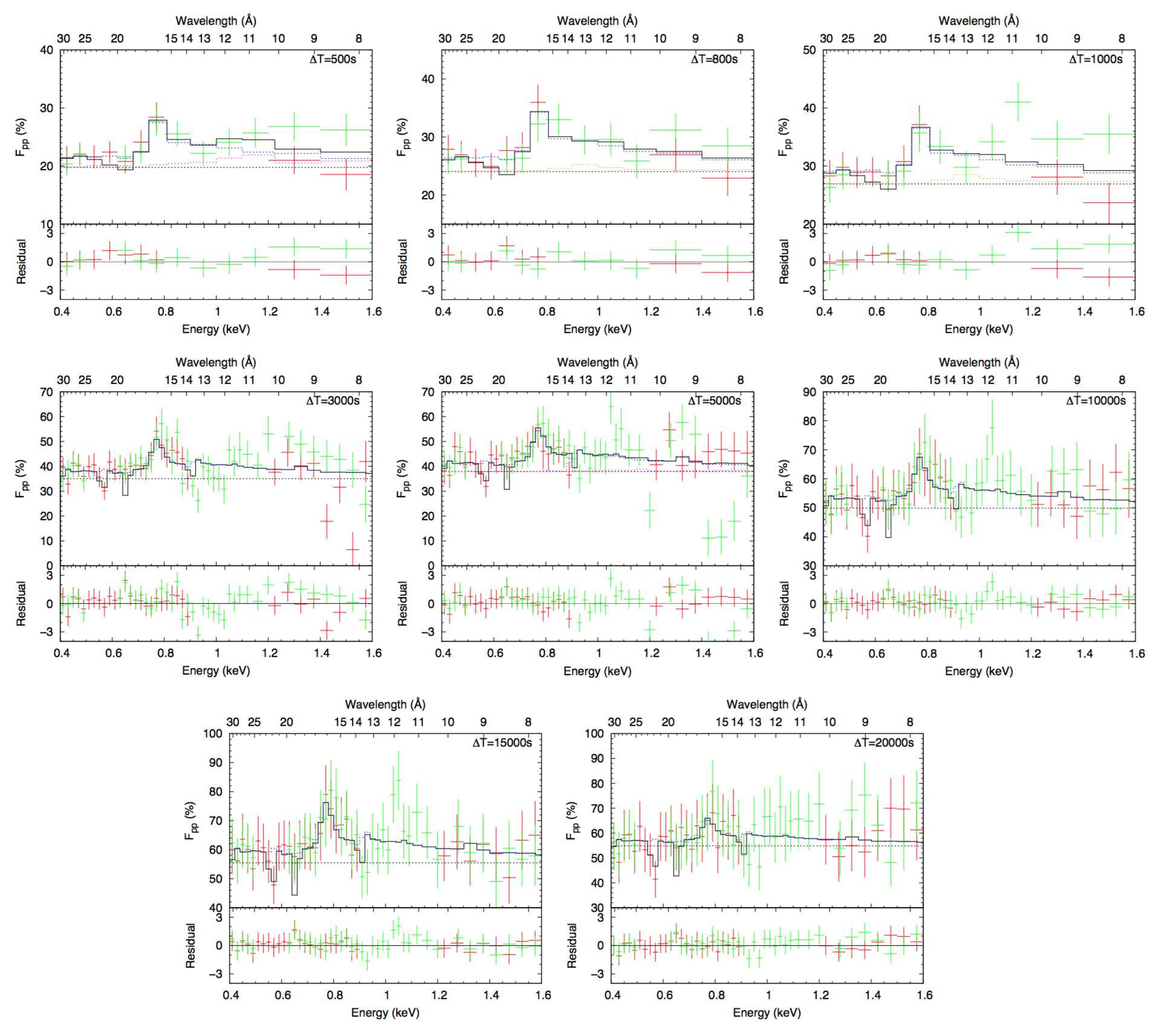}
\caption{Fitting of the $F_\mathrm{pp}$ spectra with absorption/emission models at various timescales where variability amplitude of each warm absorber and the continuum level are free.
  }\label{fig:Fpp_fitting}
\end{figure*}

Figure \ref{fig:Fpp_fitting} and Table \ref{tab:Fpp_fitting} show the fitting results of the $F_\mathrm{pp}$ spectra.
The structure of the $F_\mathrm{pp}$ spectra are similar to that of the $F_\mathrm{var}$ spectra.
WA1 shows large variability in all the examined time scales.
In particular, the variability is largest at the time scale of $\sim10000$~s.
On the other hand, WA2 and WA3 show little variability in all the time scales.


\begin{table*}
\centering
\caption{
Parameters of $F_\mathrm{pp}$ spectral fitting
  }
\label{tab:Fpp_fitting}
  \begin{threeparttable}
    \begin{tabular}{lllllllll}
      \hline
\hline
 & 500~s & 800~s & 1000~s & 3000~s & 5000~s & 10000~s & 15000~s & 20000~s \\
\hline
WA1  & $22\pm6$~\% & $30\pm7$~\% & $28\pm8$~\% & $37\pm7$~\% & $41\pm8$~\%& $41\pm10$~\% & $49\pm11$~\% & $26\pm13$~\% \\
WA2 & $0^{+11}$~\% & $7\pm18$~\% & $9\pm20$~\% & $0^{+4}$~\% & $0^{+9}$~\%& $0^{+16}$~\% & $0^{+17}$~\% & $0^{+21}$~\% \\
WA3 & $26\pm28$~\% &$0^{+29}$~\% & $0^{+95}$~\% & $0^{+33}$~\% & $10\pm40$~\% & $0^{+114}$~\% & $0^{+116}$~\% & $0^{+141}$~\%\\
\hline
Constant  & $19.8\pm0.8$~\% & $24.0\pm1.0$~\% & $26.9\pm1.1$~\% & $35.0\pm1.0$~\% & $37.9\pm1.2$~\%& $49.9\pm1.5$~\% & $55.5\pm1.7$~\% & $55\pm2$~\% \\
\hline
$\chi_\nu^2$ (dof)  & 0.62 (16) & 0.49 (16) & 1.39 (16) & 1.50 (68) & 1.44 (68)   & 0.51 (68) & 0.46 (68) & 0.44 (68)\\
\hline
      \hline
    \end{tabular}
    \begin{tablenotes}\footnotesize
      \item[$\ast$1] The fractional variability amplitude is shown.
    \end{tablenotes}
  \end{threeparttable}
\end{table*}

\subsection{Spectral variations in the whole X-ray band} \label{sec3.3}
\begin{figure*}
\includegraphics[width=150mm]{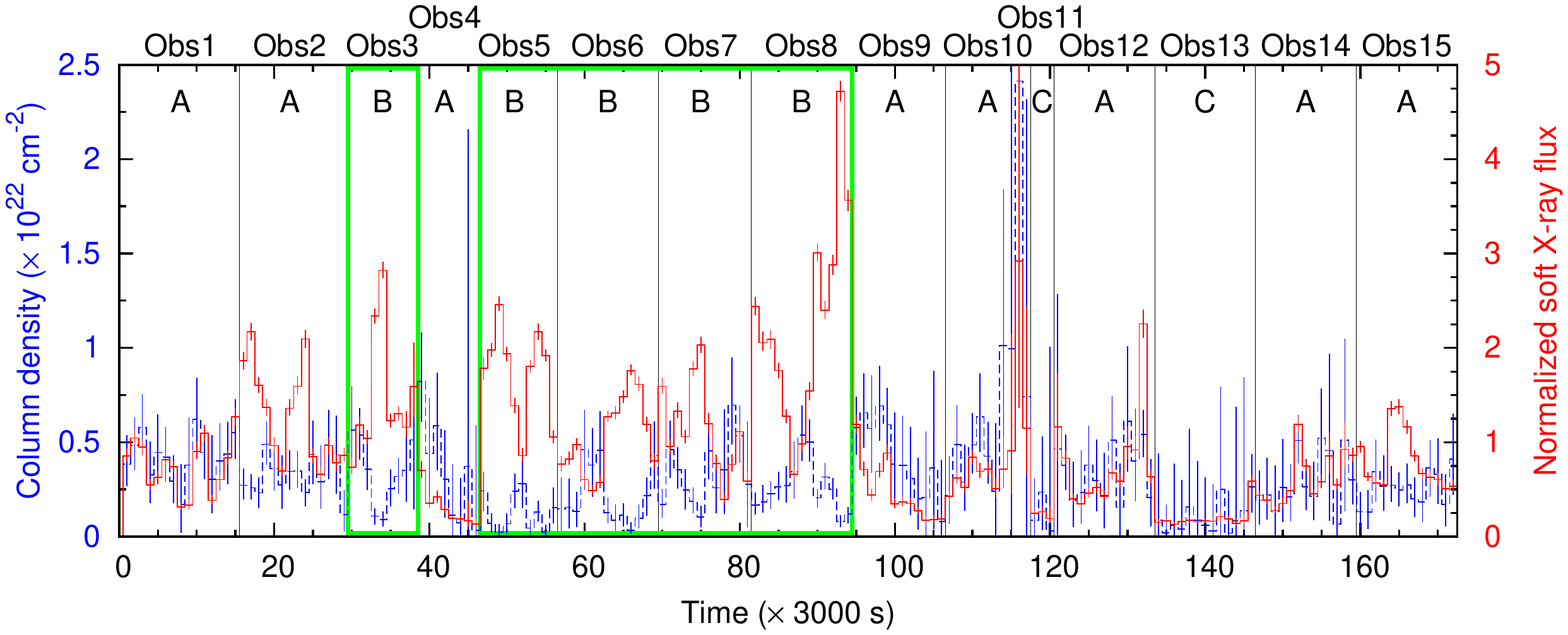}
\caption{Variations of the column density of WA1 ($=N_{\mathrm{H},1}(t)$; blue dashed) and the normalized flux observed in the soft energy band ($=C(t)$; red solid).
The time bin-width is 3000~s.
The partial covering fraction is calculated as $N_\mathrm{H}/N_\mathrm{H}^\mathrm{fixed}$.
In Obs 3 and 5 to 8 (surrounded by green lines),
anti-colleration was observed between the column density (partial covering fraction) and the observed soft X-ray flux.
See the caption of Tab.\,\ref{tab:variability} for explanation of the labels A, B, and C.
}
\label{fig:timeslice}
\end{figure*}

We have found that the observed soft X-ray flux and the parameter of WA1 show large variations in the previous subsection.
In order to investigate time variations of the parameters,
we divided the RGS spectra with a time bin-width of 3000~s, 
created 172 spectra, and fitted them with the following model,
\begin{equation}
\begin{split}
F(t)&=\mathtt{tbabs}\times\{(\mathtt{diskbb}+\mathtt{cutoffpl})\times C(t) \times \mathrm{WA1}(t) \times \mathrm{WA2} \\
&\quad \times \mathrm{WA3} +\mathtt{pexmon}+\mathrm{emission}\:\:\mathrm{lines}\}, \label{eq:eq2}
\end{split}
\end{equation}
where $C(t)$ shows variation of the contimuum normalization, or variation of the observed soft X-ray flux.
The variable parameters are only $C(t)$ and $N_{\mathrm{H},1}(t)$, where $\mathrm{WA1}(t)=\exp[-\sigma_1N_{\mathrm{H},1}(t)]$.
Variation of $N_{\mathrm{H},1}(t)$ is mathematically equivalent to variation of partial covering fraction of WA1, as shown in Eq.\,(\ref{eq:VDPC}).
All the other parameters are fixed at those in the best-fit of the time averaged spectrum (Table \ref{fitting}).
Figure \ref{fig:timeslice} shows the time-variations of $N_{\mathrm{H},1}(t)$ and $C(t)$.
We found that both the $N_{\mathrm{H},1}$ (or partial covering fraction of WA1) and the observed soft X-ray flux have large variations.
In particular, in the observational sequences of 3, 5, 6, 7, and 8 (marked with the green rectangles in Figure \ref{fig:timeslice}),
these parameters are clearly anti-correlated,
whereas in the other sequences they are not obviously correlated.
In Obs 3, 5, 6, 7, and 8, no lags are seen between the two parameters at a time scale of 3000~s.

\section{Discussion}\label{sec4}

\subsection{Origin of the observed flux/spectral variation} \label{sec4.2}
In order to explain X-ray variation of NGC 4051, several models have been proposed.
\citet{kun92} proposed a ``blob model'', where Compton-thick blobs move around the X-ray emission region, and
observed X-ray flux/spectral variations occur due to change of the number of the blobs in the line of sight.
Similarly, partial covering models have been investigated in NGC 4051,
where the observed flux/spectral variations mainly (especially in the soft X-ray energy band) result from variable partial covering fraction of the intervening absorbers (e.g.\,\citealt{pou04,hab08,ter09,lob11,iso16}).
Meanwhile, some authors argue that ionization states of the absorbers vary (\citealt{kro07,ste09,sil16}; see \S\ref{sec:otherpossibility}).
Here, we assume that the Compton-thick blobs responsible for the variation of the observed soft X-ray flux exist around the central X-ray emission region,
in addition to the ionized blobs ($=$WA1) 
responsible for the variable absorption feature.
In this manner, we introduce two partial covering layers,
one is due to the Compton-thick absorbers, and
the other is due to WA1. 
Indeed, in Eq.\,(\ref{eq:eq2}), we assume that
\begin{eqnarray}
C(t)&=&1-\alpha(t) \nonumber \\
&\simeq& 1-\alpha(t)+\alpha(t)W_\mathrm{thick},\:\:(\mathrm{when}\: E<2\,\mathrm{keV})
\end{eqnarray}
where $W_\mathrm{thick}$ indicates the Compton-thick absorber, and
\begin{equation}
\mathrm{WA1}(t)\simeq 1-\beta(t) +\beta(t)\mathrm{WA1},
\end{equation}
following Eq.\,(\ref{eq:VDPC}).
$\alpha(t)$ and $\beta(t)$ show variable partial covering fractions of
the Compton-thick absorbers and WA1, respectively.
Under these assumptions, Eq.\,(\ref{eq:eq2}) can be written as
\begin{equation}
\begin{split}
F&=\mathtt{tbabs}\times[\{(1-\alpha(t))+\alpha(t) W_\mathrm{thick}\} \times(\mathtt{diskbb}+\mathtt{cutoffpl}) \\ 
&\quad\times\{(1-\beta(t))+\beta(t) \mathrm{WA1}\}\times \mathrm{WA2}\times \mathrm{WA3} \\
&\quad +(\mathtt{pexmon}+\mathrm{emission}\:\:\mathrm{lines})]. \label{eq:eq3}
\end{split}
\end{equation}
We suppose that such a correlation exists between $\alpha(t)$ and $\beta(t)$, that, 
when the partial covering fractions increase, 
the observed soft X-ray flux gets weaker due to the Compton-thick absorbers, and
the Fe-L absorption feature gets deeper due to WA1.
When the partial covering fractions decrease, vice versa.

When the partial covering fraction of the Compton-thick blobs ($=\alpha(t)$) is variable,
observed X-ray flux in the EPIC energy band should also vary.
We created intensity-sliced spectra of the EPIC data for each observational sequence (3, 5, 6, 7, and 8) as follows:
(1) We created light curves with a bin-width of 256~s in the 0.4--12.0~keV band.
(2) We calculated all counts in each observational sequence, and determined the four intensity ranges that contain almost the same counts.
(3) From the four time-periods corresponding to the different flux levels,
we created the four intensity-sliced energy spectra in each observational sequence. 
The fitting model is Eq.(\ref{eq:eq3}).
We fixed $N_\mathrm{H,thick}$ as $1.5\times10^{24}\,\mathrm{cm}^{-2}$, which is inverse of Thomson scattering cross-section, and
$\xi$ as $10^{0.1}$, which means that $W_\mathrm{thick}$ is cold.
The velocity is fixed as $v=-660\,\mathrm{km\,s}^{-1}$, which is consistent with WA1.
The fittings were performed for each observational sequence.
We allowed the parameters of the continuum to vary among different observational sequences, but not to vary within a single sequence.
In each sequence, only $\alpha$ and $\beta$ were allowed to be free.
Namely, we assumed that, during each sequence, the intrinsic source luminosity in the fitting band (0.4--10~keV) is not variable
and the observed flux variation is only due to occultation by the partial covering blobs.

\begin{figure}
\includegraphics[width=70mm,angle=270]{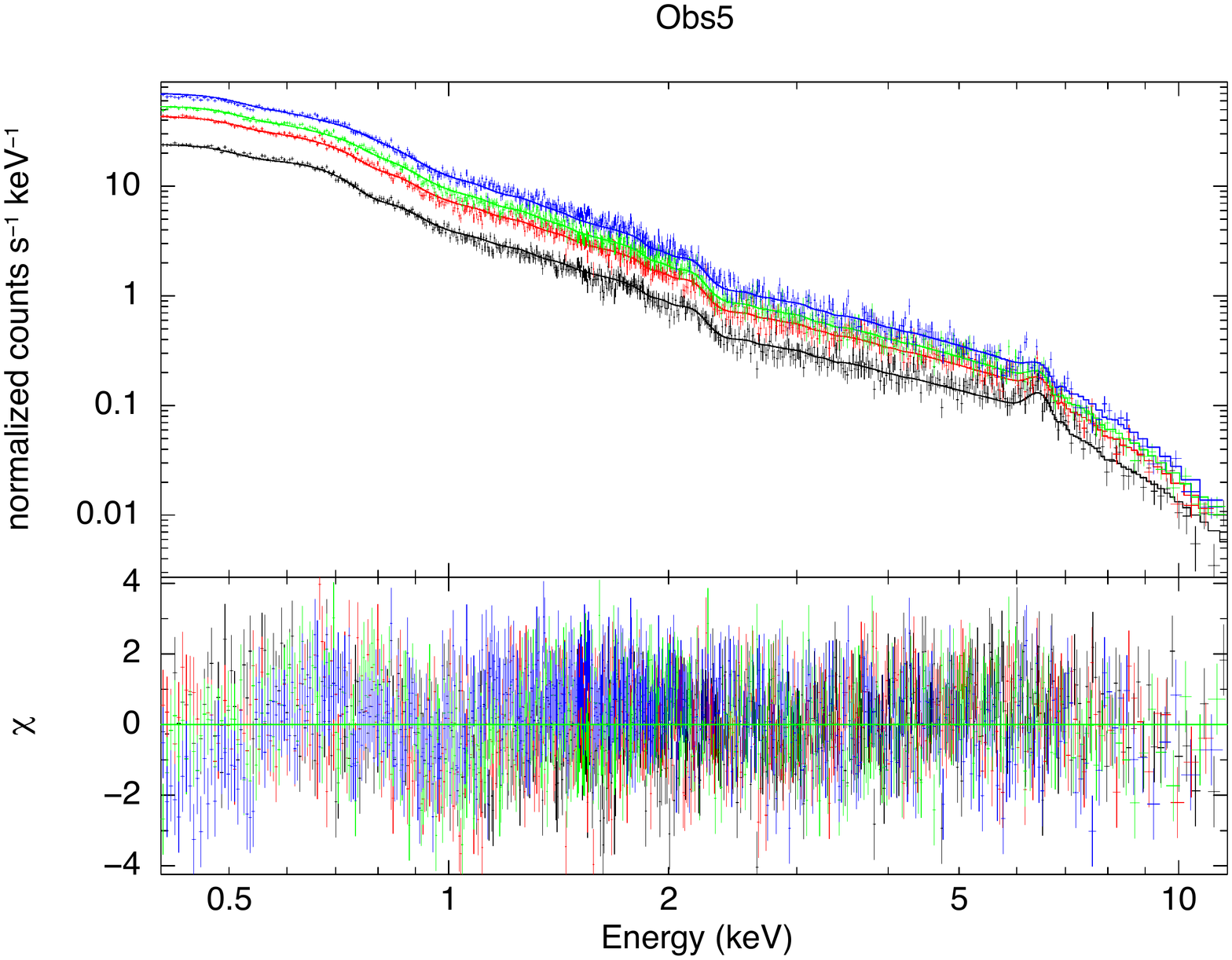}
\caption{Simultaneous fitting of the intensity-sliced spectra in Obs5.
Only the partial covering fraction of $W_\mathrm{thick}$ and the column density/the partial covering fraction of WA1 are variable.
}
\label{fig:obs5_sliced}
\end{figure}

\begin{figure}
\includegraphics[width=100mm]{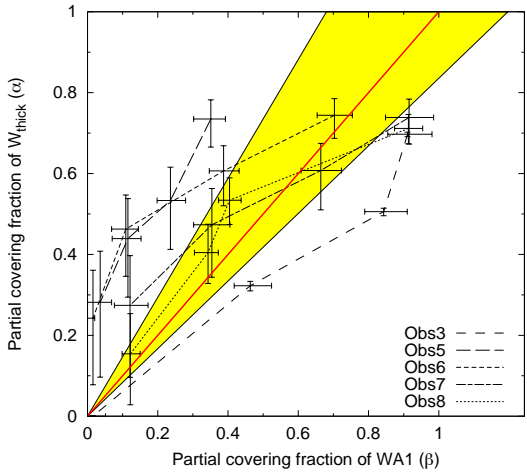}
\caption{Correlation of the partial covering fraction of $W_\mathrm{thick}$ and  WA1 when $N_\mathrm{H}^\mathrm{fixed}$ of WA1 is $6.4\times10^{21}\,\mathrm{cm}^{-2}$.
The red line shows the best-fitting, and the yellow area shows the error range.
}
\label{fig:correlation}
\end{figure}

Figure \ref{fig:obs5_sliced} shows the fitting results of Obs5.
We can explain the spectral variability in the whole 0.4--12.0~keV band with variations of only the two partial covering fractions ($\chi_\nu^2\simeq1.1$).
The fitting results of the other sequences are also reasonable ($\chi_\nu^2<1.3$).
Figure \ref{fig:correlation} shows correlation of the two partial covering fractions:
The two parameters are clearly correlated, as expected.
The red line shows the best-fit linear function with the boundary condition going through the origin ($\alpha=\beta=0$), 
and the yellow area shows the error region;
we see that $\alpha=\beta$ holds. 
The column density ($N_\mathrm{H}^\mathrm{fixed}$) of WA1 is $(6.4_{-2.1}^{+1.3})\times10^{21}\,\mathrm{cm}^{-2}$, and the average $\alpha$ is $\sim0.4$.

In Obs.\,3, 5, 6, 7, and 8, 
the observed spectral/flux variability of NGC 4051 can be explained only by change of the partial covering fraction,
whereas the intrinsic luminosity is not variable.
On the other hand, in the other observational sequences, we cannot see clear anti-corerlation between the observed flux and the partial covering fraction,
which presumably suggests that not only the partial covering fraction but also the intrinsic luminosity are variable. 
For example, in Obs.\,2, whereas the absorbers have little variations, the intrinsic luminosity variability is large, thus we can see significant variability of the observed flux.
Table \ref{tab:variability} shows variability of the partial covering fractions and the intrinsic luminosity for each observational period.
We can see the anti-correlation when the partial covering fractions vary and the intrinsic luminosity do not vary (in the situation B in Table \ref{tab:variability}).
Time-scale of the intrinsic luminosity variation is $\lesssim6000$~s ($=2$ bins in Figure \ref{fig:timeslice}), which is similar to that of the partial covering fraction.

Consequently, the spectral components and their variability are explained by the following equation:
\begin{equation}
\begin{split}
F&=\mathtt{tbabs}\times[\{(1-\alpha(t))+\alpha(t) W_\mathrm{thick}\}\times(\mathtt{diskbb}+\mathtt{cutoffpl(t)} \\ 
&\quad \times\{(1-\alpha(t))+\alpha(t) \mathrm{WA1}\})\times \mathrm{WA2}\times \mathrm{WA3} \\
&\quad +(\mathtt{pexmon}+\mathrm{emission}\:\:\mathrm{lines})]. \label{eq:result}
\end{split}
\end{equation}


\begin{table}
\centering
\caption{
Variability of the partial covering fractions and the intrinsic luminosity
  }
\label{tab:variability}
  \begin{threeparttable}
    \begin{tabular}{cccc}
      \hline
\hline
Obs. & Partial covering & Intrinsic & Observed variability\tnote{$\ast$}\\
\hline
1 & $\times$    & $\bigcirc$& A\\
2 & $\times$    & $\bigcirc$& A\\
3 & $\bigcirc$  & $\times$  & B\\
4 & $\bigcirc$  & $\bigcirc$& A\\
5 & $\bigcirc$  & $\times$  & B\\
6 & $\bigcirc$  & $\times$  & B\\
7 & $\bigcirc$  & $\times$  & B\\
8 & $\bigcirc$  & $\times$  & B\\
9 & $\bigcirc$  & $\bigcirc$& A\\
10 & $\bigcirc$ & $\bigcirc$& A\\
11 & $\times$   & $\times$  & C\\
12 & $\times$   & $\bigcirc$& A\\
13 & $\times$   & $\times$  & C\\
14 & $\times$   & $\bigcirc$& A\\
15 & $\times$   & $\bigcirc$& A\\
\hline
    \end{tabular}
\begin{tablenotes}\footnotesize
\item[$\ast$] A: The observed flux variability is mostly due to variation of the intrinsic luminosity.  \\
B: The observed flux variability is mostly due to variation of the partial covering fraction.\\
C: The observed flux is hardly variabile.
\end{tablenotes}
  \end{threeparttable}
\end{table}

\subsection{Variable double partial covering model} \label{sec:VDPC}

We have shown that ``double partial covering'' with the same partial covering fractions can explain the observed spectral/flux variability of NGC 4051 in the 0.4--10 keV.
Indeed, Eq.\,(\ref{eq:result}) is equivalent to the ``variable double partial covering (VDPC) model'', 
which was originally proposed for MCG--6--30--15 by \citet{miy12}, refined by \citet{miz14}, and confirmed for various NLS1s by \citet{iso16} and \citet{yam16}.
In the VDPC model, the commonality of the two partial covering fractions is explained by double-layer absorbers, and
the spectral variations of NLS1s are explained by the partial covering fraction of the double-layer absorbers, 
as well as independent variation of the continuum normalization.
Our results on NGC 4051 confirm validity of the VDPC model.
This double-layer absorber is very similar to the cometary one proposed in \citet{mai10}.

\subsection{Physical parameters of the warm absorber outflows}\label{sec:paraWA}

We have found that WA1 shows large variability at timescales of $\sim10$~ks.
Assuming that the warm absorber follows Kepler motion
at the distance of $r$ from the black hole, we have
\begin{equation}
\frac{r}{R_s} = 2\times10^3 
\left( \frac{\Delta T}{10^4\,\mathrm{[s]}} \right)^2
\left( \frac{D}{10\,R_s} \right)^{-2}
\left( \frac{M_\mathrm{BH}}{1.7\times10^6\,M_\odot} \right)^{-2} \label{eq:deltaT},
\end{equation}
where $R_s$ is the Schwarzschild radius,
$\Delta T$ is the timescale at which the absorber passes in front of the X-ray emission region,
$D$ is the diameter of the X-ray emission region, and
$M_\mathrm{BH}$ is the black hole mass.
WA1 shows the largest variability at $\Delta T=10^4$~s, 
which corresponds to $r\sim10^3\,R_s$,
whereas variability is seen at all the examined time scales.
WA1 and $W_\mathrm{thick}$ share the same blobs, thus the location of WA1 and $W_\mathrm{thick}$ is identical.
When the location of the blobs is $\sim 10^3\,R_s$, 
the number density and the thickness of $W_\mathrm{thick}$/WA1 are calculated as $2\times10^{12}$~cm$^{-3}$/$7\times10^{10}$~cm$^{-3}$ and $1\times10^{12}\,\mathrm{cm}$/$1\times10^{11}\,\mathrm{cm}$, respectively.
This shows that the blob is composed of a cold and dense core, and a warm and thin layer.

We cannot strongly constrain location of WA3 from the spectral variation.
From the constraint that $\Delta r \leq r$, 
WA3 presumably locates at $r\leq 3\times10^{14}\,\mathrm{cm}=6\times10^2\,R_s$, and
$n\geq 2\times10^9\,\mathrm{cm}^{-3}$.

WA2 shows little variability in both $F_\mathrm{var}$ and $F_\mathrm{pp}$ spectra,
therefore we assume that WA2 extends uniformly in the line of sight.
The number density of WA1 and WA3 are $7\times10^{10}\,\mathrm{cm}^{-3}$ and $\geq 2\times10^9\,\mathrm{cm}^{-3}$,
therefore we assume that the number density of WA2 is an order of $\sim10^{10}\,\mathrm{cm}^{-3}$.
If so, the parameters WA2 are estimated as $r\sim4\times10^{14}\,\mathrm{cm}=8\times10^2\,R_s$ and $\Delta r\sim9\times10^{11}\,\mathrm{cm}$.
Table \ref{tab:param} shows the estimated parameters of the absorbers.

\begin{table*}
\centering
\caption{
Parameters of the absorbers
  }
\label{tab:param}
    \begin{tabular}{lllllll}
      \hline
\hline
&& $n\,(\mathrm{cm}^{-3})$ & $\log\xi$ & $r\,(\mathrm{cm})$ & $\Delta r\,(\mathrm{cm})$ & $v\,(\mathrm{km\,s}^{-1})$ \\
\hline
 \multirow{2}{*}{Blobs} & Compton-thick core & $\sim2\times10^{12}$ & 0.1 & \multirow{2}{*}{$\sim10^{15}$} & $\sim1\times10^{12}$ & \multirow{2}{*}{$-660$} \\
      & Ionized layer (WA1)& $\sim7\times10^{10}$ & 1.5 &  & $\sim1\times10^{11}$ &  \\
      \hline
 \multirow{2}{*}{Line-driven disk winds} & WA2 & $\sim10^{10}$ & $2.5$ & $\sim4\times10^{14}$  & $\sim9\times10^{11}$ & $-4100$  \\
& WA3 & $\gtrsim 2\times10^{9}$ & $3.4$ & $\lesssim 3\times10^{14}$  & $\lesssim 3\times10^{14}$ & $-6100$  \\
\hline
    \end{tabular}
\end{table*}

\subsection{Comments on an alternative scenario} \label{sec:otherpossibility}
\begin{figure}
\includegraphics[width=100mm]{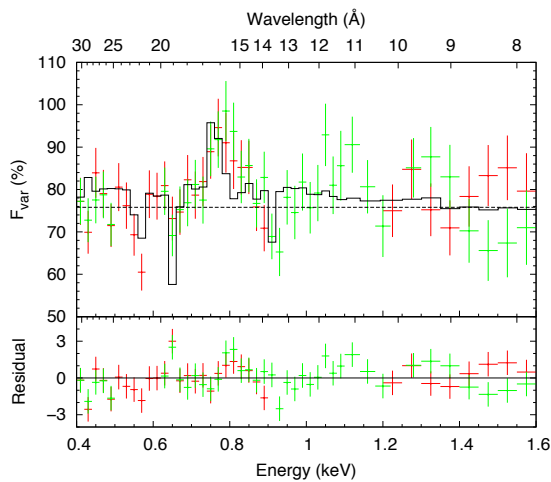}
\caption{
Fitting of the $F_\mathrm{var}$ spectra with the simulated ones
when the ionization degree of WA1 varies.
}
\label{fig:RMS_fittig_xi} 
\end{figure}

Some authors propose that observed spectral variation in NGC 4051 is 
due to variation of ionization degree, 
whereas other parameters of the warm absorbers are less variable \citep{kro07,ste09,sil16}.
In this scenario, the variable X-ray luminosity explains the ionization degree vatiaion,
such that the absorber is more ionized and transparent when the intrinsic luminosity is higher.
In order to investigate the effect of variation of the ionization degree,
we calculate the RMS spectrum when the ionization degree of WA1 is variable within $1\leq\log\xi\leq2$ by one order of magnitude \citep{kro07}.
Figure \ref{fig:RMS_fittig_xi} shows the simulated RMS spectra and the fitting results.
Whereas the centroid energy of the peak is slightly lower than that when the $N_\mathrm{H}$/$\alpha$ varies,
the fitting is reasonable ($\chi_\nu^2=1.49$ for $\mathrm{dof}=71$).

Our observational results clearly show that highly-ionized absorbers (WA2 and WA3) have little variability, whereas WA1 has large variability.
If number densities of WA2 and WA3 are sufficiently small ($n\sim10^7$\,cm$^{-3}$), 
equilibrium time scales of the absorbers is so large that we may not see ionization degree variability \citep{nic99,sil16}, and 
WA2 and WA3 are calculated to locate far from the central black hole ($r\sim10^{17}$\,cm).
However, this scenario is against the condition that the number density of WA3 is $\geq 2\times10^9\,\mathrm{cm}^{-3}$ and the distance is $\leq 3\times10^{14}\,\mathrm{cm}$ (see \S\ref{sec:paraWA}).
Furthermore, occasional independence of the observed X-ray flux and the opacity of WA1 (Figure \ref{fig:timeslice} and Table \ref{tab:variability}) has yet to be explained, since intrinsic luminosity variation should always affect ionization state of the absorber.

\subsection{Nature of the warm absorber outflows}
From the frequency- and energy-dependent time-lags,
\citet{mil10} proposes that the absorbing/reflecting clouds in NGC 4051 extend to a distance of $\simeq1.5\times10^{14}\,\mathrm{cm}$ with a covering fraction of $\gtrsim0.44$.
These physical parameters of the clouds are consistent with the partial covering blobs we found, although derived in totally different manners.
In addition, \citet{kaa14} propose that the NGC 5548 has low-ionized outflowing obscurers at a location of $\sim10^3\,R_s$, which is also similar to our partial covering blobs.
Moreover, similar clouds have been introduced in various NLS1s (see e.g.\,\citealt{mil09,iso16}).
Thus the outflowing absorbing blobs may commonly exist in the NLS1s.
The fluorescent Fe-K line of NGC 4051 is blueshifted at a velocity of $-1800\,\mathrm{km\,s}^{-1}$ (see Section \ref{sec3.1}),
which supports that the clouds not only absorb but also reflect X-rays.
Intensity of the reflection component is determined by the solid angle of the clouds,
thus non-variability of the fluorescent Fe-K line means that the solid angle of the clouds is invariable.
Those clouds in charge of the fluorescent line are out of the line of sight,
and distributed over a wide solid-angle.
Whereas the clouds in the line of sight show instantaneous variation,
solid angle of the clouds is presumably not significantly variable within each observation sequence.

WA2 and WA3 locate at $\lesssim10^3\,R_s$, close to the central X-ray source,
thus the warm absorbers are presumably launched as disk-winds,
because other mechanisms such as thermal-driven winds require extremely high temperature in this situation (e.g.\,\citealt{kin12}).
The line-driven disk wind, which is powered by radiation force due to spectral lines, is one of the plausible mechanism to produce warm absorber outflows (see e.g.\,\citealt{pro00}).
Figure 5 in \citet{nom16} shows a simulation of the geometry of the line-driven disk wind with $M_\mathrm{BH}$ of $10^6\,M_\odot$ and the Eddington ratio of $0.5$, where the outflow velocity reaches $\sim -15000$~km~s$^{-1}$.
Densities, ionization degrees, and locations of WA2 and WA3 are consistent with their simulation.
NGC 4051 is considered to have a low Eddington ratio ($\lesssim0.1$, \citealt{cze01}) than that assumed in \citet{nom16}, thus 
the outflow velocity would be slower;
this is likely to explain the observed outflow velocities of WA2 and WA3.
In this manner, we suggest that WA2 and WA3 originate in the line-driven disk winds.

\begin{figure*}
\includegraphics[width=150mm]{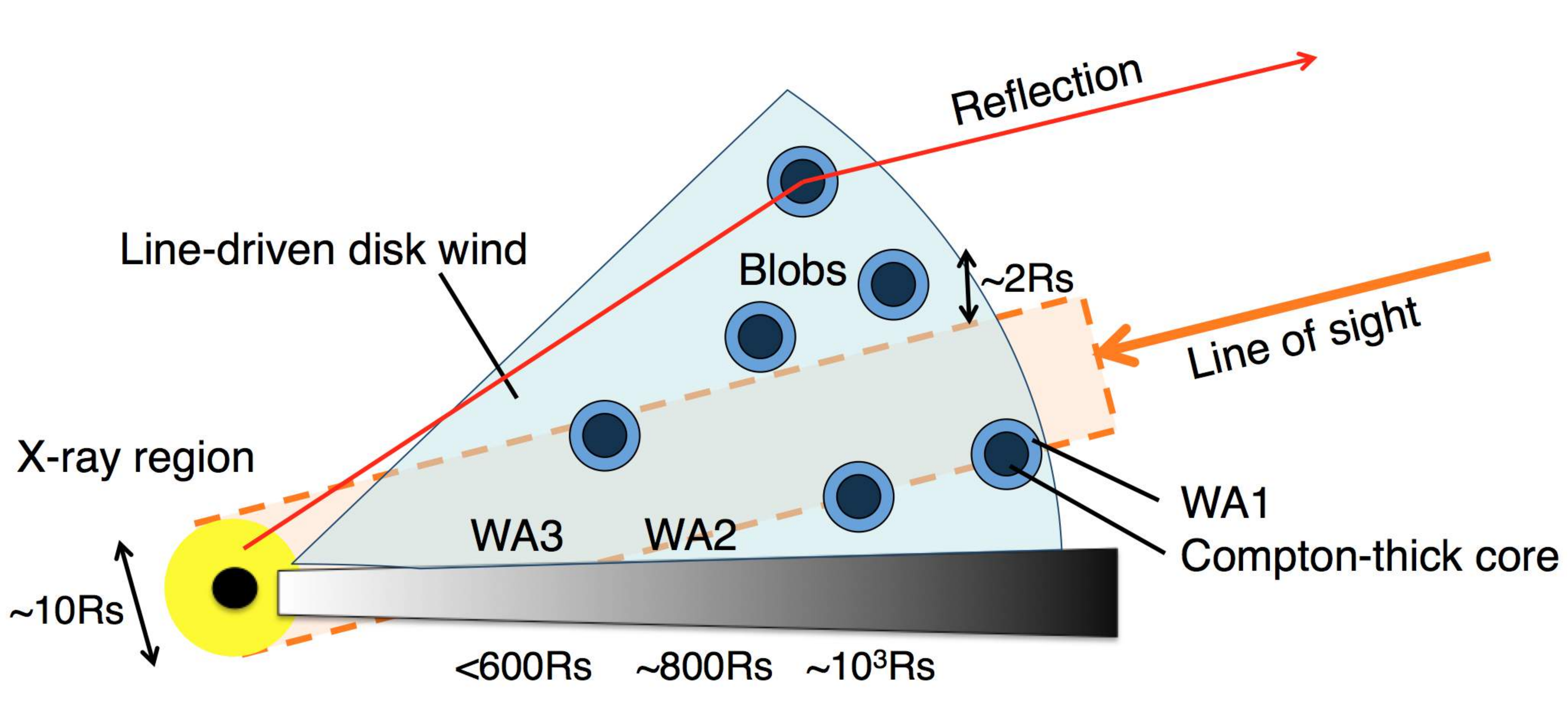}
\caption{Schematic picture of geometry of warm absorber outflows in NGC 4051
}
\label{fig:picture}
\end{figure*}

Figure \ref{fig:picture} shows a schematic picture of NGC 4051.
In summary, we have found that NGC 4051 has two types of the warm absorber outflows;
one is the line-driven disk winds, and
the other is the double-layer blobs.
The line-driven disk winds are launched within several hundred $R_s$, and
extend uniformly in the line of sight, thus they show little variability.
The double-layer absorbers are composed of the Compton-thick core and the ionized outer layer, locate at $\sim10^3\,R_s$, and partially cover the central X-ray source.
The partial covering fraction varies significantly, which is the prime origin of the observed X-ray spectral/flux variation.

\section{Conclusion}\label{sec5}

We have analyzed the XMM-Newton archival data of NLS1 NGC 4051.
X-ray energy spectra have at least three distinct warm absorber outflows; 
WA1 ($\log\xi = 1.5,\, v=-650\,\mathrm{km\,s}^{-1}$),
WA2 ($\log\xi= 2.5,\,v=-40100\,\mathrm{km\,s}^{-1}$), and
WA3 ($\log\xi= 3.4,\, v=-6100\,\mathrm{km\,s}^{-1}$).
The long enough exposure time has enabled us to calculate the RMS spectra with the RGS energy resolution for the first time.
Consequently, we have found that the RMS spectra have a sharp peak and dips, which can be explained by variable absorption features and non-variable emission lines.
The Fe-L UTA created by WA1 shows large variability at a timescale of $\sim10^4$~s,
which suggests that WA1 locates at $\sim10^3\,R_s$.
The depth of Fe-L UTA and the soft X-ray flux often show anti-correlation,
which can be explained by variable partial covering fraction of the double-layer blobs
that have the Compton-thick core responsible for the soft X-ray flux variation
and the ionized layer ($=$WA1) responsible for the Fe-L UTA,
whereas the intrinsic luminosity is hardly variable.
On the other hand, WA2 and WA3 show little variability;
they presumably extend uniformly in the line of sight, and
locate at several hundred $R_s$.
Physical parameters of WA2 and WA3 are reproduced by numerical simulation of the line-driven disk winds.
In summary,
we propose that NGC 4051 has two types of the warm absorber outflows; 
the static, high-ionized and extended line-driven disk winds, and the variable, low-ionized and clumpy double-layer blobs.

\section*{Acknowledgements}
This work is based on observations obtained with XMM-Newton, an ESA science mission with instruments and contributions directly funded by ESA Member States and NASA.
We used public data obtained through the High Energy Astrophysics Science Archive Research Center (HEASARC) at NASA/GSFC, and
software provided by HEASARC at NASA/GSFC for data reduction.
MM and KE are financially supported by the JSPS KAKENHI Grant Number JP15J07567 and JP16K05309, respectively. 

\bibliographystyle{mn}
\bibliography{mn-jour,ms}


\appendix
\section{RMS calculation}

Consider a light curve divided into $N$ time bins.
The fractional variability amplitude $F_\mathrm{var}$ is calculated to be 
\begin{equation}
F_\mathrm{var}=\frac{\sqrt{\frac{1}{N}\sum_{i=1}^{N}\left( F_i - \langle F \rangle \right)^2}}{\langle F \rangle}, \label{Fvar_def}
\end{equation}
where $F_i$ is the observed X-ray flux in the $i$th bin, and $\langle F \rangle$ is the mean value of $F_i$.

\subsection{Effect of $N_\mathrm{H}$ variation} \label{sec:A1}

Passing through an X-ray absorber,
$F_i$ is expressed as
\begin{equation}
F_i=e^{-\sigma N_{\mathrm{H},i}}F_\mathrm{int}\, ,
\end{equation}
where $\sigma$ is a cross section, $N_{\mathrm{H},i}$ is column density in the $i$th bin, and $F_\mathrm{int}$ is the intrinsic X-ray flux.

When $\tau_i=\sigma N_{\mathrm{H},i} \ll 1$ and the warm absorber is optically thin, 
\begin{equation}
F_i\simeq(1-\sigma N_{\mathrm{H},i})F_\mathrm{int}.
\end{equation}
Thus, Eq.(\ref{Fvar_def}) is calculated to be 
\begin{eqnarray}
F_\mathrm{var}&\simeq&
\frac{\sqrt{\frac{1}{N}\sum_{i=1}^{N}\left( (1-\sigma N_{\mathrm{H},i} )- \langle 1-\sigma N_\mathrm{H} \rangle \right)^2}}{\langle 1-\sigma N_\mathrm{H} \rangle} \nonumber \\
&=&\frac{\sqrt{\frac{1}{N}\sum_{i=1}^{N}\left( \sigma N_{\mathrm{H},i} - \langle \sigma N_\mathrm{H} \rangle \right)^2}}{\langle 1-\sigma N_\mathrm{H} \rangle} \nonumber \\
&=&\frac{\sqrt{\frac{1}{N}\sum_{i=1}^{N}\left( \sigma N_{\mathrm{H},i} - \langle \sigma N_\mathrm{H} \rangle \right)^2}}{\langle \sigma N_\mathrm{H} \rangle} \cdot \frac{\langle \sigma N_\mathrm{H} \rangle}{\langle 1- \sigma N_\mathrm{H} \rangle} \nonumber \\
&=& \left(\sigma N_{\mathrm{H}}\right)_\mathrm{var}\cdot\frac{ \sigma \langle  N_\mathrm{H} \rangle}{1-\sigma\langle N_\mathrm{H} \rangle} \nonumber \\
&=& \left( N_{\mathrm{H}}\right)_\mathrm{var}\cdot\frac{ \sigma \langle  N_\mathrm{H} \rangle}{1-\sigma\langle N_\mathrm{H} \rangle} \nonumber \\
&=&(N_\mathrm{H})_\mathrm{var}\cdot\frac{\langle \tau \rangle}{1- \langle \tau \rangle}. \label{Fvar_ans}
\end{eqnarray}

Eq.\,(\ref{Fvar_ans}) monotonically increases on $\sigma$ when $\langle \tau \rangle <1$.
$(N_\mathrm{H})_\mathrm{var}$ is proportional to the variable range of $N_{\mathrm{H},i}$ ($=\Delta N_\mathrm{H}$).
When $N_{\mathrm{H},i}$ monotonically varies within [$\langle N_\mathrm{H} \rangle -\Delta N_\mathrm{H}$ : $\langle  N_\mathrm{H} \rangle +\Delta N_\mathrm{H}$], $(N_\mathrm{H})_\mathrm{var}=\frac{1}{\sqrt{3}}\left(\Delta N_\mathrm{H}/\langle N_\mathrm{H} \rangle\right)$.

\subsection{Effect of partial-covering-fraction variation} \label{sec:A2}
When a warm absorber is optically thin, 
the change of $N_\mathrm{H}$ is equivalent to the change of partial covering fraction (see Eq.\ref{eq:VDPC}).
Thus, $F_i$ is expressed as
\begin{eqnarray}
F_i&=&\left(1-\alpha_i+\alpha_i\exp\left[-\sigma N_\mathrm{H}^\mathrm{fixed}\right]\right)F_\mathrm{int} \nonumber \\
&\equiv&\left(1-\alpha_i+\alpha_iW_\mathrm{f}\right)F_\mathrm{int}
\end{eqnarray}
Thus, Eq.(\ref{Fvar_def}) is calculated as
\begin{eqnarray}
F_\mathrm{var}&=&
\frac{\sqrt{\frac{1}{N}\sum_{i=1}^{N}\left\{ \left( 1-\alpha_i+\alpha_iW_\mathrm{f} \right) -  \langle 1-\alpha+\alpha W_\mathrm{f} \rangle \right\}^2}}{\langle 1-\alpha+\alpha W_\mathrm{f} \rangle} \nonumber \\
&=& \frac{\sqrt{\frac{1}{N}\sum_{i=1}^{N} (-1+W_\mathrm{f})^2(\alpha_i-\langle\alpha\rangle)^2}} { 1-\langle\alpha\rangle+\langle\alpha\rangle W_\mathrm{f} } \nonumber \\
&=& \frac{\sqrt{\frac{1}{N}\sum_{i=1}^{N}\left( \alpha_i- \langle \alpha \rangle \right)^2}}{\langle\alpha\rangle} \cdot\frac{ (1-W_\mathrm{f})\langle\alpha\rangle}{1-(1-W_\mathrm{f})\langle\alpha\rangle} \nonumber \\
&=&(\alpha)_\mathrm{var}\cdot\frac{ \left(1-\exp\left[-\sigma N_\mathrm{H}^\mathrm{fixed}\right]\right) \langle \alpha \rangle}{1-\left(1-\exp\left[-\sigma N_\mathrm{H}^\mathrm{fixed}\right]\right) \langle \alpha \rangle}\label{Fvar_ans2}
\end{eqnarray}

\subsection{Effect of an emission line} \label{sec:A3}
When a non-variable component (such as an emission line) exists, 
RMS at the energy band decreases.
Assuming that the X-ray flux consists of a variable component $A$ and an invariant component $B$,
$F_i$ is expressed as
\begin{equation}
F_i=A_i+B.
\end{equation}
Here,
\begin{eqnarray}
F_\mathrm{var}&=& \frac{\sqrt{\frac{1}{N}\sum_{i=1}^{N} \left(A_i+B-\langle A+B \rangle\right)^2 }}
{\langle A+B \rangle} \nonumber \\
&=& \frac{\sqrt{\frac{1}{N}\sum_{i=1}^{N} \left(A_i-\langle A \rangle\right)^2 }}
{\langle A \rangle +B} \nonumber \\
&=&A_\mathrm{var}\cdot\frac{\langle A\rangle}{\langle A \rangle +B}\nonumber\\
&=&A_\mathrm{var}\cdot\frac{1}{1+x},
\end{eqnarray}
where $x$ is the intensity ratio of the invariant component to the variable component ($=B/\langle A\rangle$).




\bsp	
\label{lastpage}
\end{document}